\documentclass[acmtog,article]{acmart}
\usepackage{caption}
\usepackage{booktabs} 
\usepackage{bm}
\usepackage{multirow}
\usepackage{graphicx}
\usepackage{subfigure}
\usepackage{amsmath}
\usepackage{amsfonts}
\usepackage{wrapfig}
\usepackage{rotating}
\usepackage{comment}
\usepackage{array}

\newcommand*{\bfrac}[2]{\genfrac{(}{)}{0pt}{}{#1}{#2}}

\setcitestyle{square}

\usepackage[ruled]{algorithm2e} 

\SetAlFnt{\small}
\SetAlCapFnt{\small}
\SetAlCapNameFnt{\small}
\SetAlCapHSkip{0pt}
\IncMargin{-\parindent}

\AtBeginDocument{%
  \providecommand\BibTeX{{%
    \normalfont B\kern-0.5em{\scshape i\kern-0.25em b}\kern-0.8em\TeX}}}

\copyrightyear{2020}
\acmYear{2020}

\acmJournal{TOG}
\acmYear{2020}\acmVolume{39}\acmNumber{4}\acmArticle{34}\acmMonth{7} \acmDOI{10.1145/3386569.3392491}
\setcopyright{acmlicensed}

\begin{document}

\title{Accurate Face Rig Approximation with Deep Differential Subspace Reconstruction}

\author{Steven L. Song}
\authornote{Authors contributed equally.}
\email{stevens@blueskystudios.com}
\affiliation{%
  \institution{Blue Sky Studios}
  \streetaddress{1 American Ln}
  \city{Greenwich}
  \state{CT}
  \postcode{06831}
}
\author{Weiqi Shi}
\authornotemark[1]
\email{weiqi.shi@yale.edu}
\affiliation{%
  \institution{Yale University}
  \city{New Haven}
  \state{CT}
  \postcode{06520}
}

\author{Michael Reed}
\email{reed@blueskystudios.com}
\affiliation{%
  \institution{Blue Sky Studios}
  \streetaddress{1 American Ln}
  \city{Greenwich}
  \state{CT}
  \postcode{06831}
}

\begin{abstract}

To be suitable for film-quality animation, rigs for character deformation must fulfill a broad set of requirements. They must be able to create highly stylized deformation, allow a wide variety of controls to permit artistic freedom, and accurately reflect the design intent. Facial deformation is especially challenging due to its nonlinearity with respect to the animation controls and its additional precision requirements, which often leads to highly complex face rigs that are not generalizable to other characters. This lack of generality creates a need for approximation methods that encode the deformation in simpler structures. We propose a rig approximation method that addresses these issues by learning localized shape information in differential coordinates and, separately, a subspace for mesh reconstruction. The use of differential coordinates produces a smooth distribution of errors in the resulting deformed surface, while the learned subspace provides constraints that reduce the low frequency error in the reconstruction. Our method can reconstruct both face and body deformations with high fidelity and does not require a set of well-posed animation examples, as we demonstrate with a variety of production characters.

\end{abstract}



\begin{teaserfigure}
\centering
\subfigure{
\begin{minipage}[b]{0.21\textwidth}
\includegraphics[width=1\textwidth]{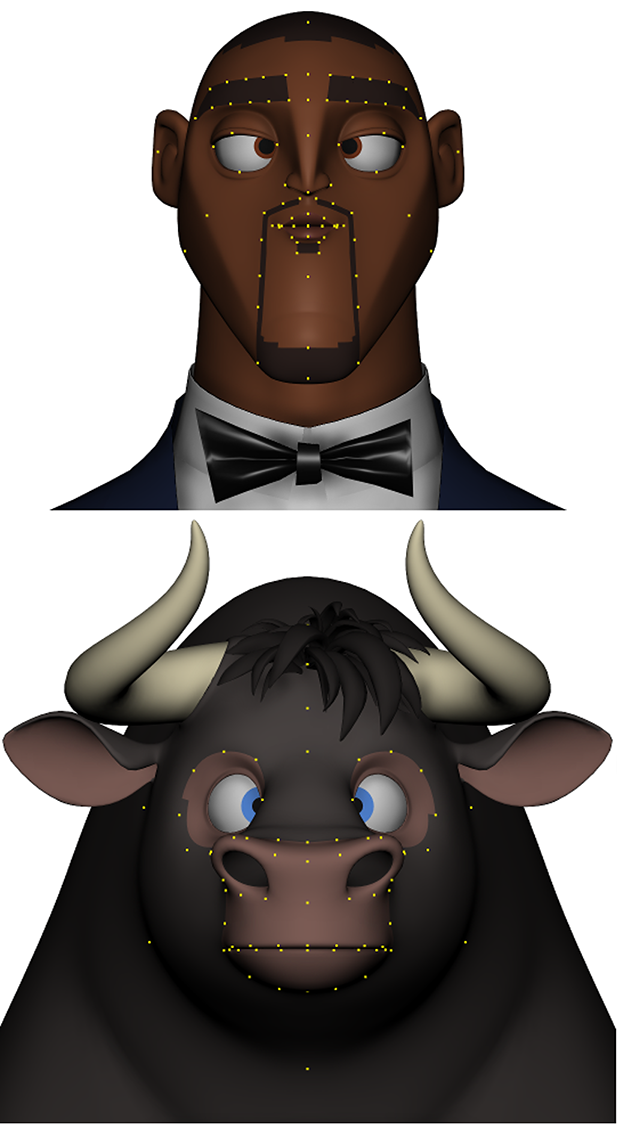} 
\vspace{-2em}
\caption*{Rest Pose}
\end{minipage}
}\hspace{0em}
\subfigure{
\begin{minipage}[b]{0.21\textwidth}
\includegraphics[width=1\textwidth]{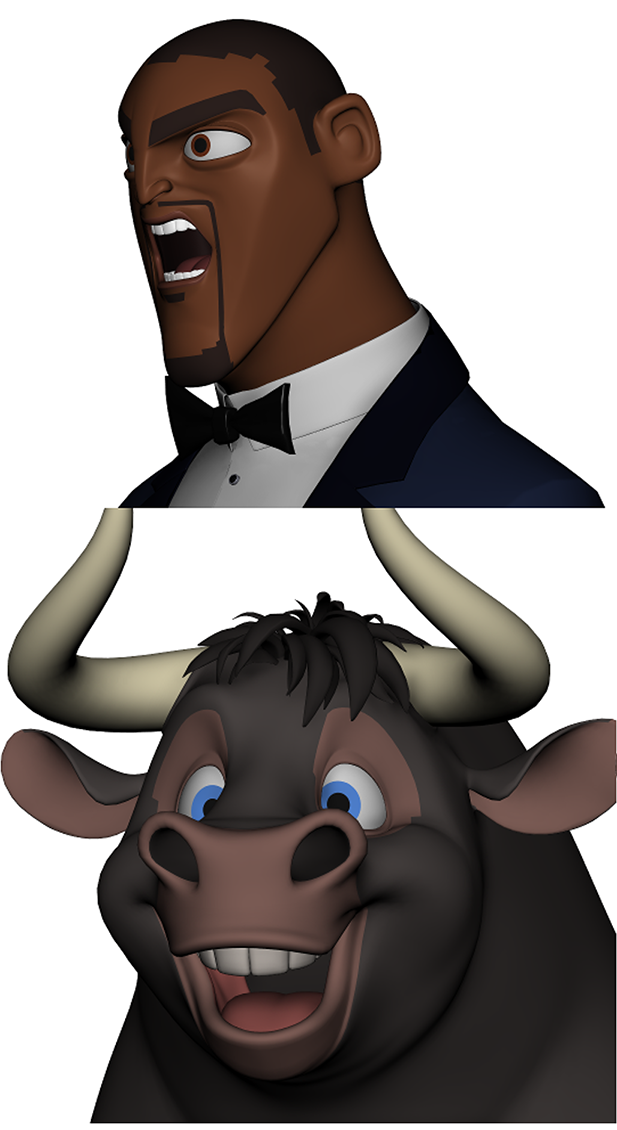}
\vspace{-2em}
\caption*{Ground Truth}
\end{minipage}
}\hspace{0em}
\subfigure{
\begin{minipage}[b]{0.21\textwidth}
\includegraphics[width=1\textwidth]{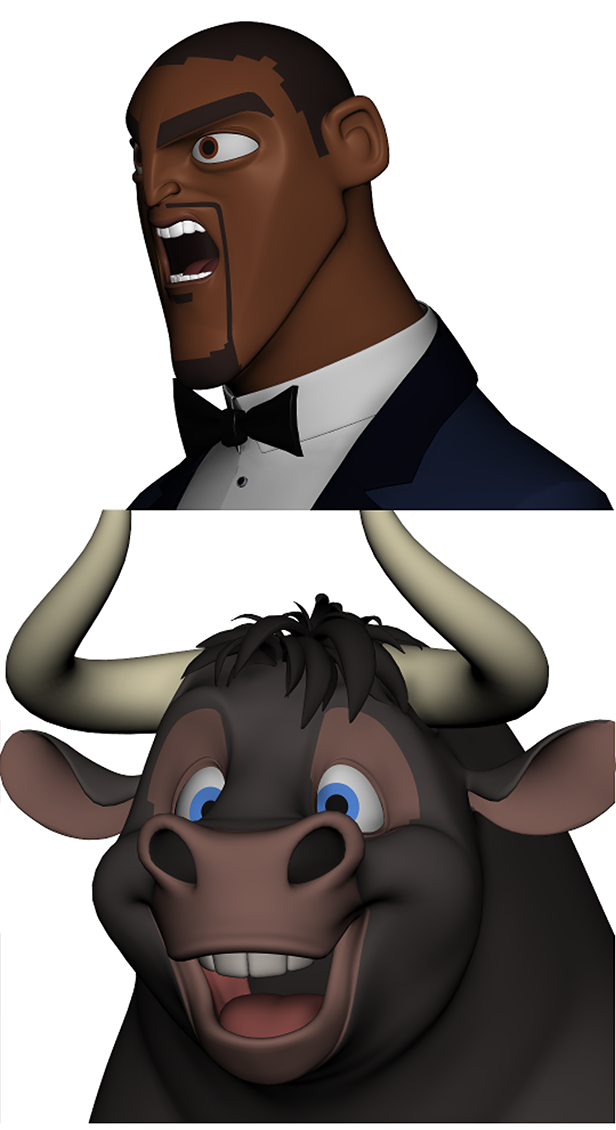}
\vspace{-2em}
\caption*{Our Method}
\end{minipage}
}\hspace{0em}
\subfigure{
\begin{minipage}[b]{0.21\textwidth}
\includegraphics[width=1\textwidth]{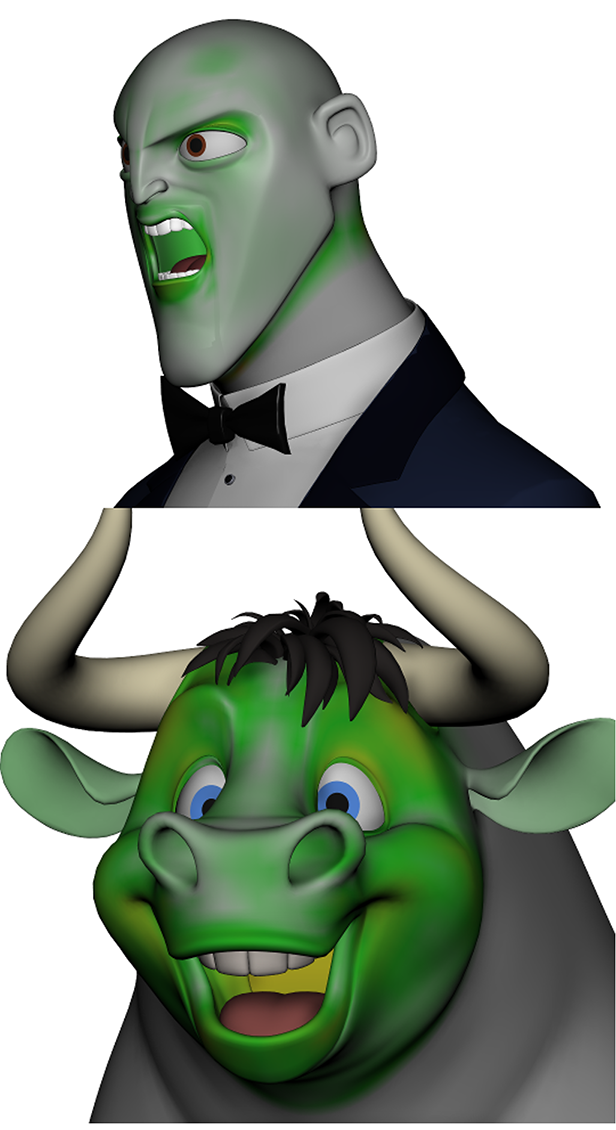} 
\vspace{-2em}
\caption*{Difference}
\end{minipage}
}\hspace{0em}
\subfigure{
\begin{minipage}[b]{0.06\textwidth}
\includegraphics[width=1\textwidth]{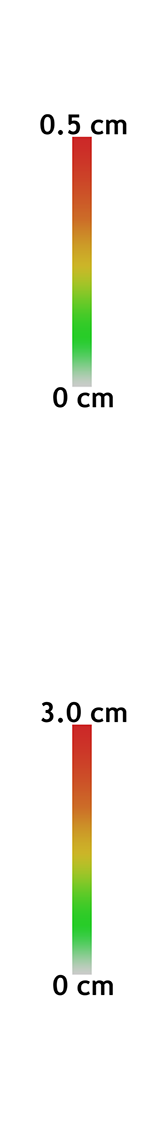} 
\vspace{-2em}
\caption*{ }
\end{minipage}
}
\vspace{-2em}
  \caption{Our rig approximation method learns localized shape information in differential coordinates and, separately, a subspace for mesh reconstruction.}
  \label{fig:teaser}
\end{teaserfigure}

\maketitle

\section{Introduction}

Film-quality character rigs rely on a complex hierarchy of procedural deformers, driven by a large number of animation controls, that map to the deformation of the vertices of a character's surface mesh. Because the characters are subject to high aesthetic standards, and the rigs are the primary means by which the animators interact with them, the rigs themselves have strict performance requirements: the character's skin must behave predictably and precisely over the entire range of control, which for animated characters can be extreme because of the caricatured design and motion. 

Rigs for facial animation typically have much more complex behavior than body rigs, and require additional precision due to their importance in conveying the most crucial aspects of communication and expression. To offer artistic freedom, the face rig is usually a complex structure containing a large number of numerical controls. Unlike the joint-based controls commonly used for a character's body, these numerical controls are globally defined and cooperatively influence the transformation of each vertex, making facial deformation highly nonlinear and expensive to compute.

In production it's often desirable to reuse the same rig behavior for different purposes in different environments. For example, transferring the rig to a simulation application for crowd simulation, to a game engine for VR production, or to a renderer for render-time manipulation. Unfortunately it's often not viable to take the original rig to other packages because a visually-matching reimplementation is required per deformer per package.  Similarly, simulation-based rigs (e.g. muscle systems) provide complex behavior that is desirable in many production situations, but their lack of interactive response discourages their adoption. These issues can be addressed by a rig approximation method if it has the following characteristics: simple universal structure, high accuracy and good performance. A neural network approach automatically meets the first requirement, as the same network can approximate varying non-linear functions with different sets of weights. Neural networks can also provide benefits with batch evaluation. For example crowd characters, which can often reuse the same nonlinear deformation with different scaling factors, can be batch evaluated if driven by a neural network. Much of the work in this area -- on moving from the typical rig deformer ``stack'' to a neural representation -- has focused on run-time performance e.g. \cite{bailey2018fast}.


 In contrast, our work directly addresses the importance of accuracy as experienced in the film production environment. In this paper we introduce a new learning-based solution to accurately capture facial deformation for characters using differential coordinates and a network architecture designed for that space. Similar to other work, we assume that the deformation has both a linear and a nonlinear component that can be separated. The linear deformation is not the focus of this paper since its contribution to facial deformation is limited and many linear skinning solutions have been proposed \cite{kavan2008geometric, kavan2005spherical}. Instead, we focus on learning the nonlinear component, which applies equally well to both face and body rig approximation, as we show in our results.

At run-time our method takes as input animation controls defined as a set of artist-level rig parameters, and computes the deformation as vertex displacements from the rest pose. During the offline training process, we use vectorized features generated from rig parameters, and labels are differential coordinates calculated from the localized nonlinear deformation of the original rig. The differential coordinates have the advantages of a sparse mesh representation and embedded neighbor vertex information, which contribute to the learning of local surface deformation. However, the transformation between coordinates is ill-conditioned and non-invertible, and so we introduce a separate subspace to improve the conditioning of the reconstruction. This subspace is determined by artist-specified ``anchor points'', selected from the original mesh at features that are significant to the character's expressive ability. Our method conducts separate subspace training to learn how these anchor points deform using a split network structure. 

We qualitatively and quantitatively evaluate our method on multiple production-quality facial rigs. Experimental results show our method can predict accurate facial deformation with minimal visual difference from the ground truth. We show our method extends to body deformation where it compares favorably with existing solutions. Additionally, we show how using anchor points improves the reconstruction by reducing the low frequency error introduced in the differential training.

\section{Related Work}
\subsection{Skinning and Rigging}
Skinning techniques can be roughly divided into physics-based \cite{kim2017data, si2014realistic}, example-based \cite{mukai2016efficient, loper2015smpl}, and geometry-based methods. We focus here on geometry-based solutions due to their computational efficiency and simplicity. One of the most widely used techniques is linear blend skinning (LBS) \cite{magnenat1988joint}, where a weighted sum of the skeleton's bone transformations is applied to each vertex. Advances in this technique include dual quaternion skinning (DQS) \cite{kavan2008geometric}, spherical blend skinning \cite{kavan2005spherical} and optimized centers of rotation skinning \cite{le2016real}.  Although these methods are computationally efficient for computing linear deformation, they do not handle nonlinear behaviors such as muscle bulging and twisting effects. Improving on this, Merry et al. \shortcite{merry2006animation} and Wang et al. \shortcite{wang2002multi} introduce more degrees of freedom for each bone transformation through additional skin weights, which can be acquired by fitting example poses. Other approaches designed to address these issues include pose space deformation \cite{lewis2000pose, sloan2001shape}, cage deformation \cite{jacobson2011bounded, joshi2007harmonic, ju2005mean, lipman2008green}, joint-based deformers \cite{kavan2012elasticity}, delta mush \cite{mancewicz2014delta, le2019direct} and virtual/helper joints methods \cite{kavan2009automatic, mukai2015building, mukai2016efficient}. Wang et al. \cite{wang2007real} introduce a rotational regression model to capture nonlinear skinning deformation, which optimizes the deformation of all vertices simultaneously using the Laplace equation. An iterative optimization \cite{sorkine2007rigid} is proposed to approximate nonlinear deformation by alternating surface smoothing and local deformation. All of these methods require additional computational cost for nonlinear components and are primarily focused on body deformation, leaving facial deformation largely unaddressed.

\subsection{Facial Rig and Deformation}
In contrast to body rigs that are defined by bones and joints, facial rigs often include hundreds of animation controls represented by numerical values which control the nonlinear transformation of each vertex. These animation controls are globally defined and widely used in blendshapes \cite{lewis2010direct, lewis2014practice} to achieve realistic facial animation for production. Prior work focused on editing data-driven facial animation \cite{deng2006animating, joshi2006learning} or providing intuitive control \cite{lau2009face, lewis2010direct}. Li et al. \cite{li2010example} successfully transfer controller semantics and expression dynamics from a generic template to the target model using blendshape optimization in gradient space. Weise et al. \cite{weise2011realtime} present a high-quality performance-driven facial animation system for capturing facial expressions and creating a digital avatar in real-time. A blendshape system that allows efficient anatomical and biomechanical facial muscle simulation is proposed in \cite{cong2016art}.

\subsection{Learning-based Deformation}
There has been increasing interest in using learning-based solutions to replace traditional deformation algorithms. Previous work such as \cite{lewis2000pose} utilize a support vector machine to learn mesh deformation given a set of poses. \cite{tan2018variational, tan2018mesh} propose mesh-based autoencoders to learn deformation from a latent space. Based on their work, \cite{gao2018automatic} put forward a solution to transfer shape deformation between characters with different topologies using a generative adversarial network. Luo et al. \shortcite{luo2018nnwarp} propose a deep neural network solution to approximate nonlinear elastic deformation, combining this with simulated linear elastic deformation to achieve better results. Liu et al.\shortcite{liu2019neuroskinning} use graph convolutional networks to predict the skin weight distribution for each vertex, resulting in a trained network that can be applied to different characters given their mesh data and rigs. Relevant to our work is \cite{bailey2018fast}, where multiple neural networks are used to approximate the rig's nonlinear deformation components under the assumption that each vertex is associated with a single bone. For each bone, they train a network to predict the offset of each associated vertex. Three unaddressed issues that motivate our work are: (1) the deformation of a vertex is often influenced by multiple bones, with no single bone as the prominent influence, (2) the deformation can be determined by numeric controls (as in face rigs) and (3) associating bones with disjoint sets of vertices can introduce discontinuities at set boundaries.

\subsection{Subspace Deformation and Model Reduction}
Subspace model reduction techniques are commonly used to solve nonlinear deformation in real-time applications. Instead of evaluating the complete mesh, subspace models compute the deformation of a low dimensional embedding on the fly and project it back to the entire space.  Subspace deformation was originally used in early simulation work \cite{pentland1989good}, which uses a subspace spanned by the low-frequency linear vibration modes to represent the deformation. To augment the linear model and handle non-linearities, Krysl et al. \shortcite{krysl2001dimensional} propose the empirical eigenvectors subspaces using principal component analysis (PCA) for finite element models.  Summer et al. \shortcite{sumner2007embedded} use graph structure to represent deformations as a collection of affine transformations for shape manipulation. An et al. \shortcite{an2008optimizing} introduces subspace forces and Jacobians associated with subspace deformations for simulation. Barbi\v{c} et al. \shortcite{barbivc2005real} observe that the reduced internal forces with linear materials are cubic polynomials in reduced coordinates, which could be precomputed for efficient implicit Newmark subspace integration. For deformation-related model reduction, Barbi\v{c} et al. \shortcite{barbivc2012interactive} propose a method for interactive editing and design of deformable object animations by minimizing the force residual objective.
Wang et al. \shortcite{wang2015linear} design linear deformation subspaces by minimizing a quadratic deformation energy to efficiently unify linear blend skinning and generalized barycentric coordinates.  Building on these works, a recent hyper-reduced scheme \cite{brandt2018hyper} uses two subspaces to achieve real-time simulation, one for constraint projections in the preprocessing stage and the other for vertex positions in real-time. Close to our work is Meyer et al. \shortcite{meyer2007key}, who propose the Key-Point Subspace Acceleration (KPSA) and caching to accelerate the posing of deformable facial models. The idea of using key points for reconstruction is analogous to the anchor points in our case. However, their method, like other subspace techniques, relies on high quality animation prior examples to compute the embedding of the subspace. 

Compared with previous work, the advantages of our method are: (1) it can reconstruct both face and body deformation with high accuracy,  (2) it can take different types of animation controls as input, (3) it does not require a particular set of well-posed animation priors and (4) it provides a simple universal structure for cross-platform real-time evaluation.

For the rest of the paper, we first review the preliminaries of differential coordinates in Section 3.1. We then describe our training pipeline (Section 3.2), including the vectorization from input animation controls, the acquisition of nonlinear deformation from existing poses, network structures and reconstruction. We introduce the implementation details in Section 3.3, and we describe our experiments, evaluate the training results, compare with existing solutions in Section 4. Finally, Section 5 discusses limitations and future work.

\begin{figure*}
  \centering
  \vspace{-1em}
  \includegraphics[width=0.95\linewidth,scale=0.18]{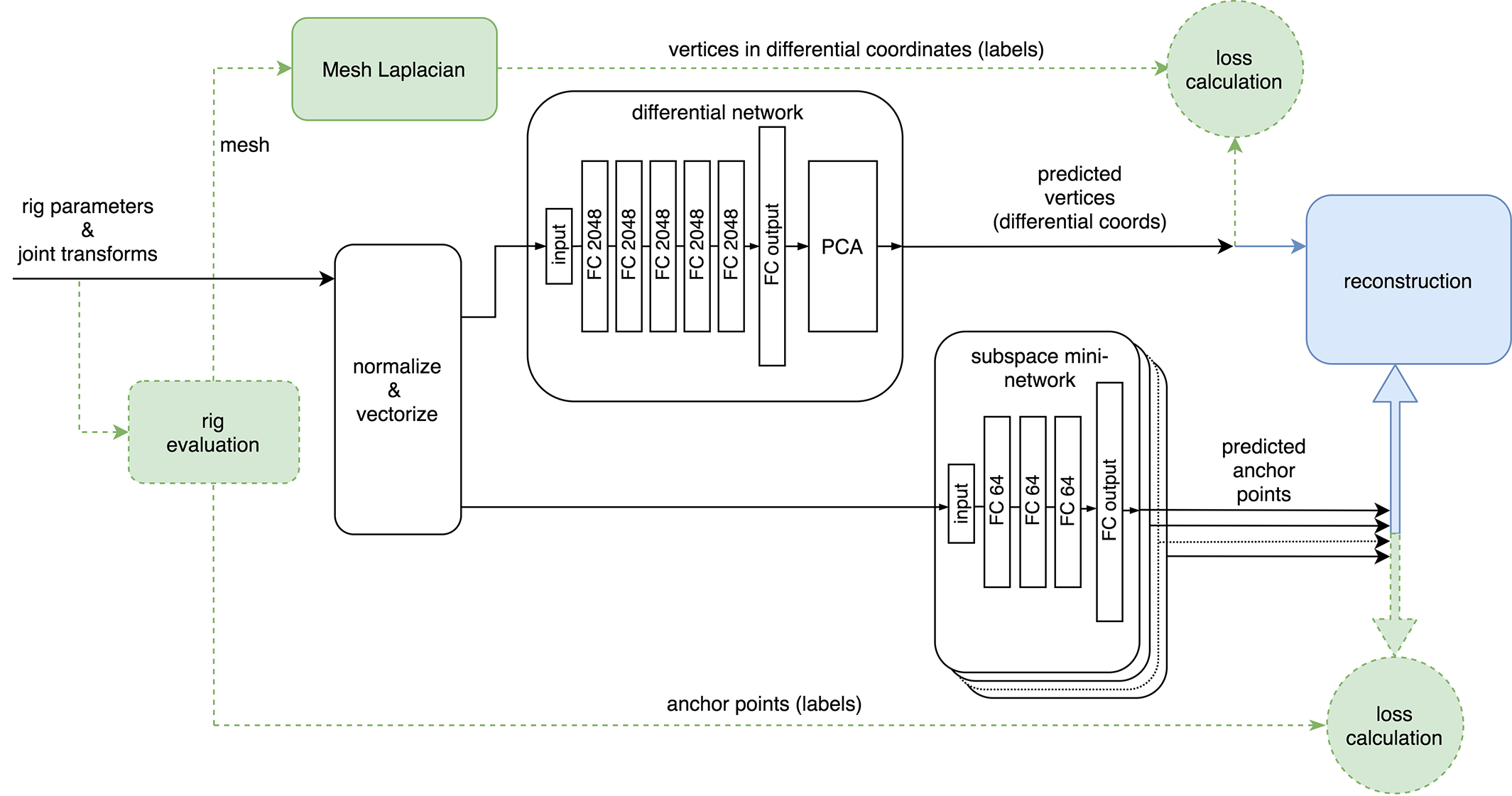}
  \caption{\label{fig:pipeline} Our method takes rig parameters and the corresponding joint transforms as input and predicts the nonlinear deformation of the mesh vertices (in differential coordinates) and the set of anchor points (in cartesian space). Green pathways are for network training, blue pathways for prediction.
}
\end{figure*}

\section{Method}

Our model approximates the nonlinear deformation in a character rig. The linear deformation can be simply represented with linear blend skinning, so it's not our focus here. For a given mesh in rest pose, our model takes animation controls defined by a set of rig parameters as inputs, and outputs the non-linear deformation of the mesh. Fig.~\ref{fig:pipeline} shows our training pipeline. To process the training data, we first vectorize the input rig parameters and extract the nonlinear deformation represented by vertex displacement from the corresponding deformed mesh. Then we convert the nonlinear deformation into differential coordinates ($\delta$ space), where we learn localized shape information and map the rig controls to it. However, we cannot directly reconstruct the mesh surface from differential coordinates since the transformation is ill-conditioned. We conduct a separate subspace learning on a group of anchor points selected from the original mesh, for which we learn deformation in local coordinates and use them as constraints for reconstruction.

\subsection{Preliminary}
Let $ \bm{M} \in \{\bm{V}, \bm{E}\} $ be a mesh with $n$ vertices, $\bm{V} \in \mathbb{R}^{n \times 3}$. Each vertex $\bm{v}_i \in \bm{V}$ is represented using absolute Cartesian coordinates and $\bm{E}$ represents the set of edges. The Laplacian operator $\bm{L}$ is defined \cite{sorkine2005laplacian} as:

\begin{equation}\label{eq1}
\bm{L} = \bm{I}- \bm{D}^{-1}\bm{A}
\end{equation}

where $\bm{A}$ is a (0, 1) adjacent matrix of size $n \times n$ that indicates the connectivity of vertex pairs in the mesh with $\bm{A}_{ij} = 1$ if $(i,j) \in \bm{E}$. $\bm{D}$ is a diagonal matrix of size $n \times n$ representing the degree $d_i$ of each vertex. Applying the Laplacian operator $\bm{L}$ to the vertices transforms the mesh into delta space, where each vertex $\bm{v}_i$ is represented as $\bm{\delta}_i$. The differential coordinate of each vertex represents the difference between the vertex itself and the center of mass of its immediate neighbors ($A_i$ denotes the neighborhood set of vertex $\bm{v}_i \in \bm{V}$):

\begin{equation}\label{eq2}
\begin{gathered}
\bm{L}\bm{V} = \bm{\delta} \\
\bm{v}_i - \frac{1}{d_i}\sum_{j\in A_i}^{}\bm{v}_j =  \delta_i
\end{gathered}
\end{equation}

It's more convenient to use the symmetrical version of $\bm{L}$, denoted by $\bm{L_s} = \bm{D}\bm{L} = \bm{D} - \bm{A}$, giving:

\begin{equation}\label{eq3}
\bm{L_s}\bm{V} = \bm{D}\bm{\delta}
\end{equation}

Compared to the Cartesian coordinates, where only the spatial location of each vertex is provided, the differential coordinates carry information about the local shape of the surface and the orientation of local details. It preserves local surface detail and captures the irregular shape of the surface. Transferring mesh deformation data into differential space leads to a sparse representation, which also contributes to the learning process. Intuitively, if a surface patch is deformed uniformly, the differential representation of the deformation will have zero values for all vertices except for the boundaries.

Given the Laplacian operator and differential coordinates, we now consider how to reconstruct mesh surface. Note the matrix $\bm{L_s}$ is singular and has a non-trivial zero eigenvector because the sum of all its rows is 0. Therefore, we cannot directly invert the matrix for reconstruction, but can add constraints to the matrix to make it full rank. We introduce the subspace $\bm{P}$, which is constructed by a set of anchor points from $\bm{V}$. The dimension of the subspace is much smaller than the original mesh. The index matrix of the anchor points $I(\bm{P})$ is appended at the end of the Laplacian matrix $\bm{L_s}$. Correspondingly, we append the Cartesian coordinates of anchor points $\bm{V}(\bm{P})$ to the differential coordinates of the full mesh to make it solvable:

\begin{equation}\label{eq4}
\widetilde{\bm{L}}\bm{V} = \bfrac{\bm{L_s}}{\bm{\omega} I(\bm{P})} \bm{V} = \bfrac{\bm{D}\bm{\delta}}{\bm{\omega} \bm{V}(\bm{P})} = \bm{\widetilde{\delta}}
\end{equation}

$\widetilde{\bm{L}}$ is the full-rank matrix with anchor points appended to the original Laplacian matrix. $\bm{\omega}$ is the weight matrix for the anchor points, which can be used to stress the importance of each anchor points.  Given the full rank matrix $\widetilde{\bm{L}}$  and $\bm{\widetilde{\delta}}$, we can solve the following equation:

\begin{equation}\label{eq5}
(\widetilde{\bm{L}}^T\widetilde{\bm{L}})\bm{V} = \widetilde{\bm{L}}^T\bm{\widetilde{\delta}}
\end{equation}

Applying the Laplacian operator to a mesh is analogous to obtaining the second spatial derivatives. The eigenvectors of $\bm{L}$ are cosine basis functions of the Fourier transform, and the associated eigenvalues are squares of the frequencies \cite{zhang2010spectral}. We demonstrate that for a small error $\bm{\epsilon}$ introduced in differential coordinates, the high frequency component of $\bm{\epsilon}$ is dampened when converted back to Cartesian space. This leads to a smoother distribution of the error, which is much less noticeable in the reconstructed surface. 

Since $\bm{L_s}$ is symmetric positive semi-definite, it has an orthogonal eigenbasis $\bm{E} = \{\bm{e_1},\bm{e_2},...\bm{e_n}\}$, with corresponding eigenvalues $0<\lambda_1 \leq \lambda_2 \leq \lambda_3 \leq...\leq \lambda_n$. (For this analysis, we assume $\bm{L_s}$ is non-singular by adding one anchor)

\begin{equation}\label{eq6}
\begin{aligned}
\bm{L_s}\bm{V^\prime} &= \bm{D}(\bm{\delta} + \bm{\epsilon}) \\
\bm{V^\prime} &= \bm{L_s}^{-1}\bm{D}(\bm{\delta} + \bm{\epsilon}) \\
\bm{V^\prime} &= \bm{V} + \bm{L_s}^{-1}\bm{D}\bm{\epsilon}
\end{aligned}
\end{equation}

We denote $\bm{D}\bm{\epsilon}$ as $\bm{\epsilon^\prime}$ and decompose it in basis $\bm{E}$
\begin{equation}\label{eq7}
\bm{\epsilon}^\prime = c_1\bm{e_1}+c_2\bm{e_2}+...c_n\bm{e_n} 
\end{equation}

Notice that $\bm{L_s}^{-1}$ shares the same eigenvectors and its corresponding eigenvalues are inversed. We have

\begin{equation}\label{eq8}
{\bm{L_s}}^{-1}\bm{\epsilon}^\prime =\frac{1}{\lambda_1}c_1\bm{e_1}+\frac{1}{\lambda_2}c_2\bm{e_2}+...\frac{1}{\lambda_n}c_n\bm{e_n}
\end{equation}

Since $\lambda_1$ is small and $\lambda_n$ is large, the inverse of the eigenvalues amplifies the low frequency eigenvector $\bm{e_1}$ and dampens the high frequency one $\bm{e_n}$. In this way, the high-frequency errors in the differential coordinates are reduced. This is desirable for mesh deformation as localized high frequency errors are much more noticeable. To reduce the amplification of low-frequency error, we increase the number of anchor points, which improves the conditioning of the Laplacian matrix by increasing the smallest singular value. Therefore, we can decrease both the low and high-frequency errors when the mesh surface is reconstructed.

\subsection{Pipeline}

\subsubsection{Input Features} The rig parameters cannot be directly used for training because they are in different representations and scales. Therefore, we need to first create feature vectors from the given rig parameters. Without loss of generality, we assume that facial rigs include joint controls $\bm{J}$ and numerical controls $\bm{C}$. For the joint controls, we use the transformation matrix $\bm{M}_{\bm{J}_{i}} = [\bm{X}_{\bm{J}_{i}}, \bm{t}_{\bm{J}_{i}}]$ of each joint $\bm{J}_{i}$ as input, where $\bm{X}_i \in \mathbb{R}^{3 \times 3}$ is the rotation/scale matrix and $\bm{t}_i \in \mathbb{R}^{3}$ is the normalized translation value. We vectorize and concatenate all the joint controls so that we have $\bm{J} = \{\bm{J}_{1},...\bm{J}_{i},...\bm{J}_{j}\},\bm{J}_{i} \in \mathbb{R}^{12}$. For the numerical controls, we define the input features as the concatenation of the normalized numerical value of each attribute, $\bm{C} = \{C_{1},...C_{i},...C_{c}\},C_{i} \in \mathbb{R}^1$, where $C_{i}$ represents each control attribute. Then we concatenate all the joint and numerical controls as our input feature $\bm{F}$, whose dimension is $12j+c$. We normalize all the translation values together, but every single numerical control attribute is normalized independently since they are on different scales.

\begin{equation}\label{eq9}
\bm{F} = Concat(||^{j}_{i=1}\bm{J}_{i},||^{c}_{i=1}C_{i})
\end{equation}

\begin{figure}
  \centering
  \hspace{-1em}
  \includegraphics[width=1\linewidth,scale=0.25]{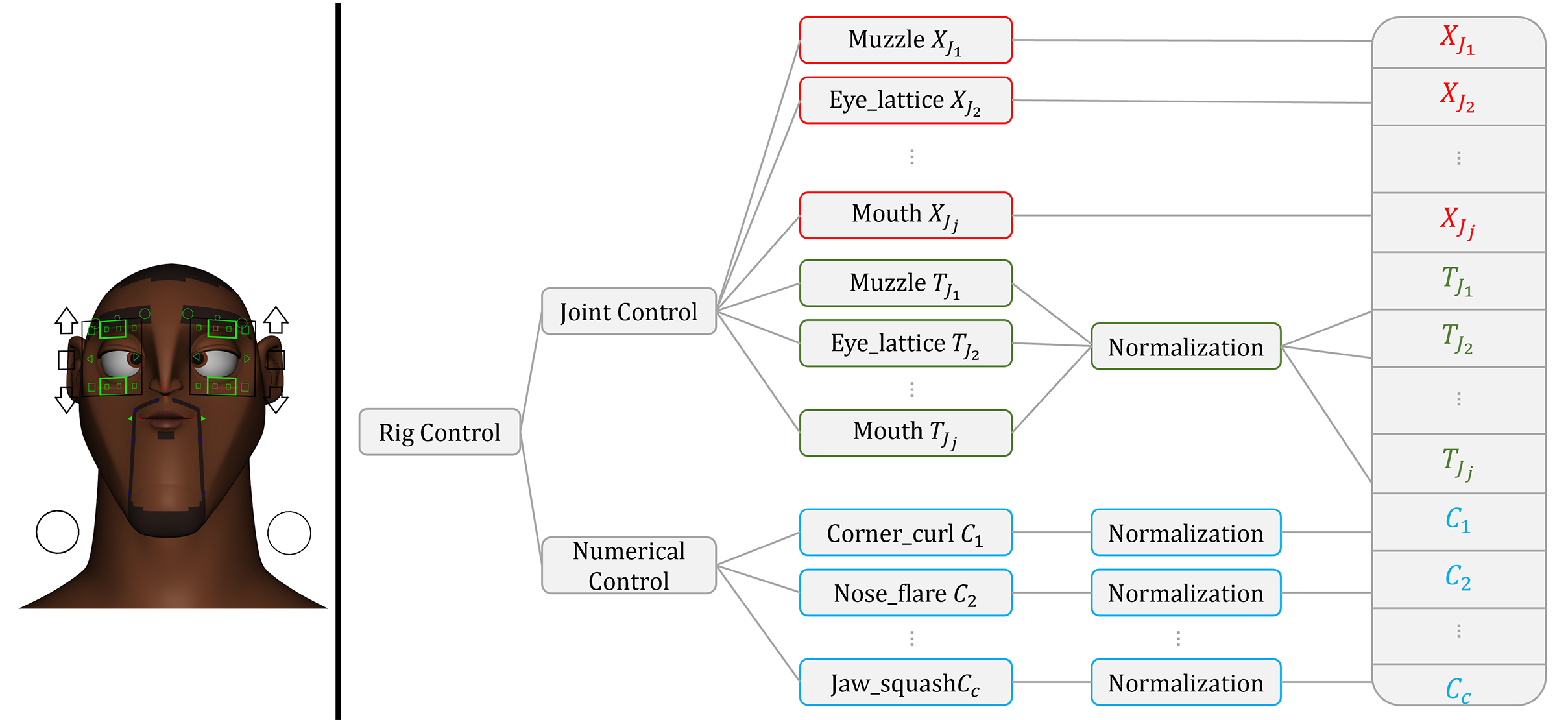}
  \vspace{-1em}
  \caption{\label{fig:control}An example for rig controls and vectorization. Only joint controls are shown on the character.}
  \vspace{-1em}
\end{figure}

To generate the training data, we randomly and independently sample each rig control using truncated Gaussian distribution within a set range. The range of each control is defined so that it reasonably covers the possible range of animation, similar to the method used by \cite{bailey2018fast}. We do not limit our training data to well-animated poses because (1) they require human labor and thus are expensive to generate, and (2) using randomly generated poses can cover a large range of motion and more dynamic deformations, which can improve the generalization of our model.

\subsubsection{Nonlinear Deformation} We use the nonlinear deformation as our training labels, which can be computed from the deformed mesh. We assume a mesh in rest pose $\bm{V}$  and its deformation $\widetilde{\bm{V}}$ is defined by a set of rig parameters. We also assume $\widetilde{\bm{V}}$ and $\bm{V}$ maintain the same topology. The vertex $\bm{v}_i \in \bm{V}$ and $\widetilde{\bm{v}}_i \in \widetilde{\bm{V}}$ are defined in local Cartesian coordinates. We have the following equation:

\begin{equation}\label{nlcompute}
\widetilde{\bm{v}}_i = \bm{T}_i(\bm{v}_i + \bm{v}_{i,nl})
\end{equation}

where $\bm{v}_{i,nl}$ is the vertex displacement in local space caused by the nonlinear deformation. $\bm{T}_i$ is the linear transformation for vertex $\bm{v}_i$ which can be computed from the transformation matrix of the joint controls.

\begin{equation}\label{explicit}
\bm{T}_i = \sum_{k=1}^{\bm{J}(\bm{v}_i)}\omega_{k}\bm{M}_{\bm{J}_{k}}(\bm{M}^{o}_{\bm{J}_k})^{-1}
\end{equation}

$\bm{J}(\bm{v}_i)$ represents the joint controls that have influence on the vertex $\bm{v}_i$. $\bm{M}_{\bm{J}_{k}}$ denotes the transformation matrix for joint $\bm{J}_{k}$ and $\bm{M}^{o}_{\bm{J}_k}$ is its transformation matrix at rest pose. $\omega_{k}$ is the weight for the joint. We assume the rig as a black box, so we don't have $\bm{M}^{o}_{\bm{J}_k}$ and $\omega_k$ available. For general purposes, we use an implicit method to calculate $\bm{T}_i$. Given equation \ref{nlcompute}, we perturb $\bm{v}_{i,nl}$ by moving one unit for every direction along XYZ coordinates and observe the vertex displacement produced by the rig. Then we can use the vertex displacement to calculate $\bm{T}_i$. With the following equations:

\begin{equation}\label{eq12}
\begin{aligned}
\widetilde{\bm{v}}_{i}^{\prime}&= \bm{T}_i\bm{v}_i\\
\widetilde{\bm{v}}_{i,x}&= \bm{T}_i(\bm{v}_i + (1,0,0,0)^{T}) \\
\widetilde{\bm{v}}_{i,y}&= \bm{T}_i(\bm{v}_i+  (0,1,0,0)^{T}) \\
\widetilde{\bm{v}}_{i,z}&= \bm{T}_i(\bm{v}_i+  (0,0,1,0)^{T}) \\
\widetilde{\bm{v}}_{null}&= \bm{T}_i(0,0,0,1)^{T} \\
\end{aligned}
\end{equation}

By subtracting the first equation from the following ones, we have:

\begin{equation}\label{eq13}
\bm{T}_i = (\widetilde{\bm{v}}_{i,x}-\widetilde{\bm{v}}_{i}^{\prime},\quad \widetilde{\bm{v}}_{i,y}-\widetilde{\bm{v}}_{i}^{\prime}, \quad \widetilde{\bm{v}}_{i,z}-\widetilde{\bm{v}}_{i}^{\prime}, \quad \widetilde{\bm{v}}_{null})
\end{equation}

$\bm{T}_i$ can be substituted into equation \ref{nlcompute} to calculate the nonlinear deformation with given rig input:

\begin{equation}\label{result}
\bm{v}_{i,nl} = \bm{T}_i^{-1}\widetilde{\bm{v}}_i - \bm{v}_i
\end{equation}

Our goal is to learn the nonlinear deformation from given rig parameters by minimizing the per-vertex distance between our results and the ground truth.

\subsubsection{Differential Network} The differential network takes the vectorized features as input and outputs the vertex displacement corresponding to the nonlinear deformation in differential coordinates. This network has 5 fully connected layers with 2048 units, each followed by a Relu activation layer. Similar to \cite{laine2017production,bailey2018fast}, we apply PCA at the end of the network by multiplying the projection matrix with the output. We precompute the projection matrix on the entire training set. The training data can be constructed as a matrix $\bm{M} \in \mathbb{R}^{3|V| \times m}$ where $|V|$ is the vertex count and $m$ is the dimension for all training poses. The purpose of PCA is to project the network output back to a lower dimension which helps the network converge. We determine the number of principal components as a fixed percentage of the number of mesh vertices, which is simple to implement in practice (we evaluate the influence of different percentage on training in Section 4.1). Alternatively the PC number can be selected by choosing the most significant basis vectors such that the reprojection error of the training set is below a defined threshold.


For the loss function, a simple choice would be the regression loss such as the Euclidean distance between the predicted vertex displacement and the ground truth. However, it is known that an L2 loss function tends to blur the prediction results \cite{isola2017image, liu2019neuroskinning}. The mesh deformation for character animation is smooth and continuous, which implies the differential representation has small values. Our training data is generated by random sampling the rig parameters, but this also means the training data contains outliers that would never appear in real animation and which appear in delta space as large values. L2 loss is more sensitive to outliers due to the consideration of the squared differences. In our case, L2 loss tends to adjust the network to fit and minimize those outlier vertices, which leads to higher errors for other vertices. On the other hand, using L1 loss reduces the influence of outliers and produces better result. Therefore we use the L1 loss for the differential network.

\subsubsection{Subspace Network} The subspace network takes the vectorized features as input and outputs the nonlinear deformation of selected anchor points in local Cartesian coordinates for reconstruction. Previously, Chen et al. \shortcite{chen2005algebraic} and Sorkine et al. \shortcite{sorkine2005geometry} use greedy heuristic methods to select anchor points. They treat all the vertices in the mesh equally and iteratively select the vertex based on the largest geodesic distance between the approximated shape and the original mesh. However, these algorithms do not fit in our situation because of the different contributions of vertices to the facial animation. We pay more attention to the important facial features, such as eyes and mouth, rather than nose, ears or the scalp. In general face rigs define the controls on those areas to constrain the deformation. Therefore, we use the rig as reference to select anchor points and make sure that they are well-distributed and proportional to the density of the rig controls. Based on our observation, the training performance and reconstruction results do not depend on the specific anchor point selection as long as the major deformable facial features are covered. We also note that the number of anchor points contributes to the accuracy of reconstruction; we evaluate that in Section 4.2.

The subspace network consists of a set of mini-networks, each of which corresponds to a single anchor point and outputs its deformation in $\mathbb{R}^{3}$. For the input of each mini-network, we perform a dimension reduction technique similar to that used in \cite{bailey2018fast}, where each network takes as input a subset of the vectorized features corresponding to the rig controls that deform the anchor point. However, the difference between our method and Bailey et al. is that we perform the split training on the anchor points instead of the entire mesh, and so we avoid the discontinuity issue. We apply this technique because only a small subset of all rig controls influence a certain anchor point. We collect the related rig controls by perturbing all the controls individually and recording which anchor produces deformation. This process is repeated with 100 random example poses and with large perturbations to ensure that controls affecting the anchor are identified. Each mini-network includes 3 fully connected layers with 64 units, each followed by a Relu activation layer. For the loss function, we use L2 loss for the network as the subnetwork is trained on Cartesian coordinates, which don't encode mesh information in a way that accentuates outliers. We use multiple mini-networks instead of a single network because there is no direct spatial relationship between the anchor points and there is low correlation between their deformation. In practice, we found this structure has better training performance compared with the single network due to the reduced dimension. 

\subsubsection{Reconstruction} We perform reconstruction using the full-rank Laplacian matrix $\widetilde{\bm{L}}$, which is constructed by appending the indices of anchor points at the end of the original Laplacian matrix $\bm{L}$. Notice $\widetilde{\bm{L}}$ does not vary with input rig parameters and only depends on the selected anchor points. According to equation \ref{eq5}, we can apply Cholesky factorization on $\widetilde{\bm{L}}^T\widetilde{\bm{L}}$ to get the upper-triangular sparse matrix $\bm{R}$:

\begin{equation}\label{eq15}
\widetilde{\bm{L}}^T\widetilde{\bm{L}} = \bm{R}^T\bm{R}
\end{equation}

We only need to compute the factorization once with only the mesh topology information and the matrix $\bm{R}$ can be reused whenever rig parameters change. Now we can easily solve the equation \ref{eq4} and reconstruct the mesh surface using back substitution. We concatenate the results from the differential and subspace network to get $\bm{\widetilde{\delta}}$ and use it in the following equation:

\begin{equation}\label{eq16}
\bm{R}^T\bm{R}\bm{V_{nl}} = \widetilde{\bm{L}}^T\bm{\widetilde{\delta}}
\end{equation}

Since $\bm{R}$ is a triangular matrix, we can efficiently reconstruct the nonlinear deformation $\bm{V_{nl}}$ with back substitution, which makes it possible to run the reconstruction at an interactive speed with frequently updated results from the networks. 

We use uniform Laplacian instead of the cotangent Laplacian because the latter changes as the mesh deforms, requiring expensive recomputation for every pose. With uniform Laplacian the factorization only needs to happen once, and the reconstruction is done with 2 back substitutions, which are very fast.

\subsection{Implementation Details}

For both the differential and subspace networks, we set the batch size as 128 and choose a SGD solver for optimization with the initial learning rate as 0.1 and the learning rate decay as $10^{-6}$ (SGD outperforms Adam in our case). We train 10000 epochs for both the network, which takes 3.5 hours for the differential network on an NVIDIA GeForce GTX 2080 GPU, and less than 1 hour for the subspace network.

\section{Evaluation}
\begin{figure*}[tb]
\centering
\subfigure{
\begin{minipage}[b]{0.1\textwidth}
\includegraphics[width=1\textwidth]{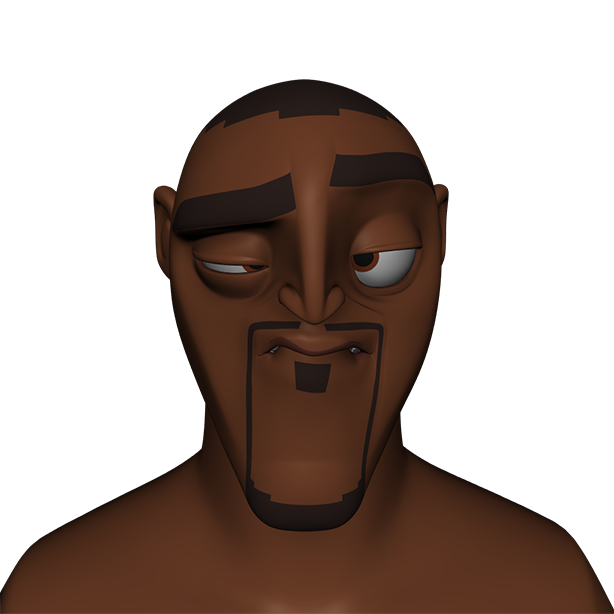} \\
\includegraphics[width=1\textwidth]{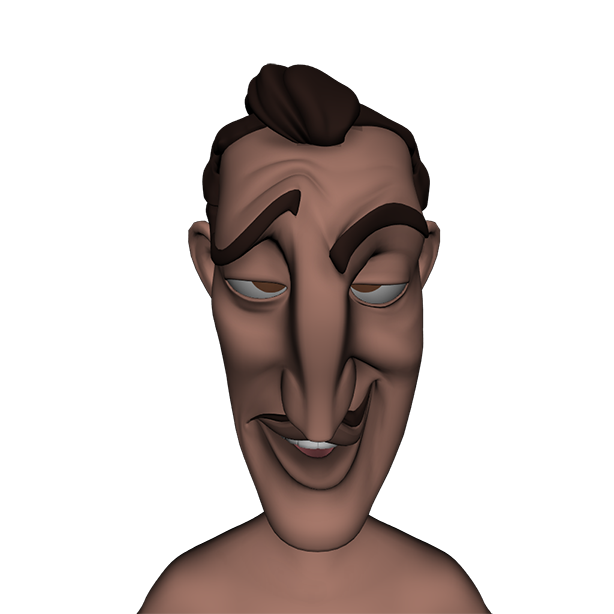} \\
\includegraphics[width=1\textwidth]{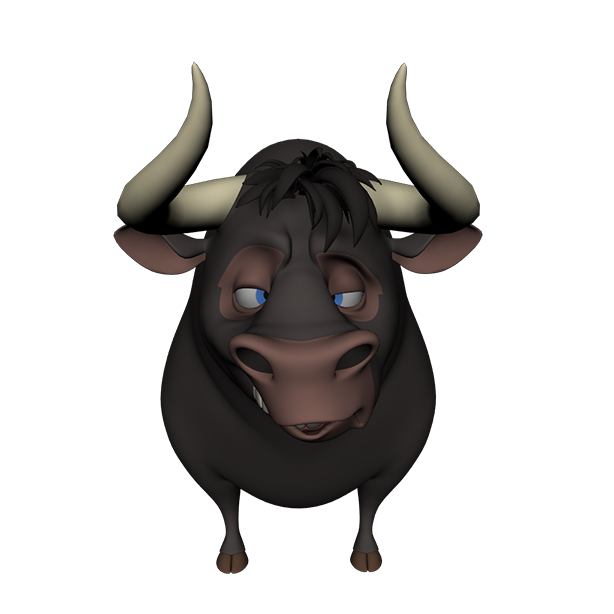}
\vspace{-2em}
\end{minipage}
}\hspace{2mm}
\subfigure{
\begin{minipage}[b]{0.1\textwidth}
\includegraphics[width=1\textwidth]{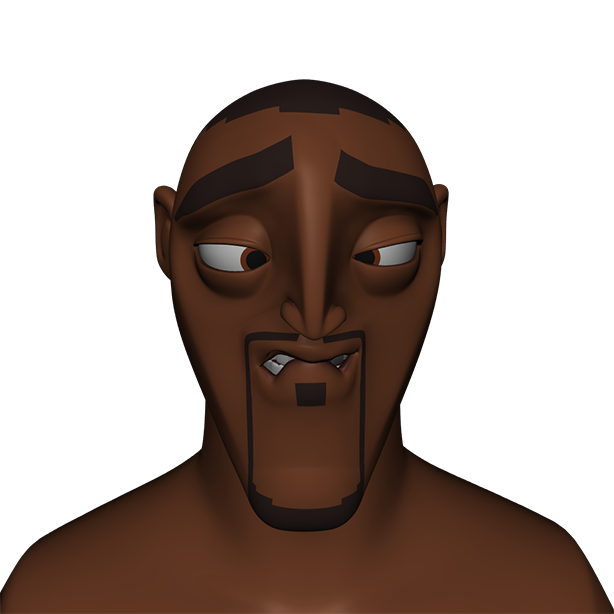} \\
\includegraphics[width=1\textwidth]{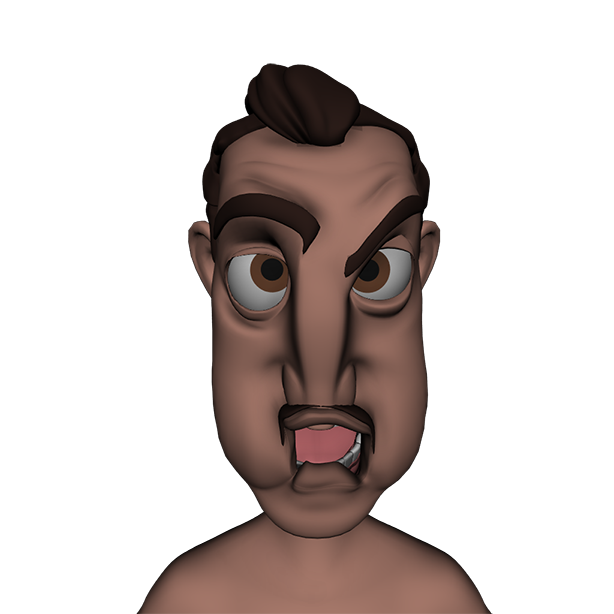} \\
\includegraphics[width=1\textwidth]{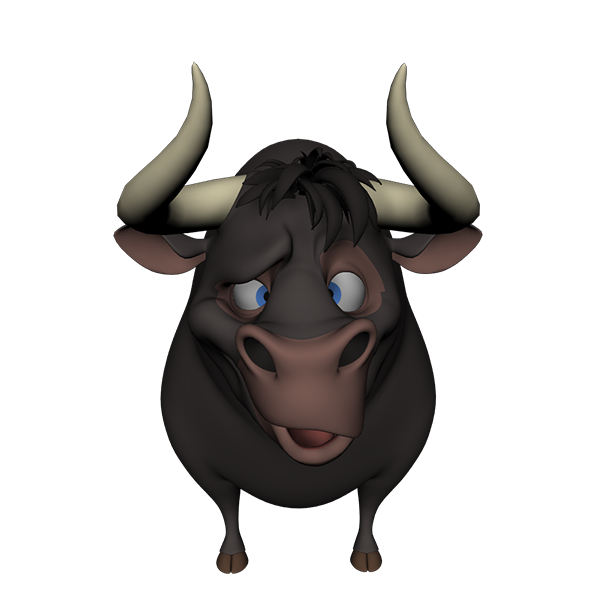}
\vspace{-2em}
\end{minipage}
}\hspace{2mm}
\subfigure{
\begin{minipage}[b]{0.1\textwidth}
\includegraphics[width=1\textwidth]{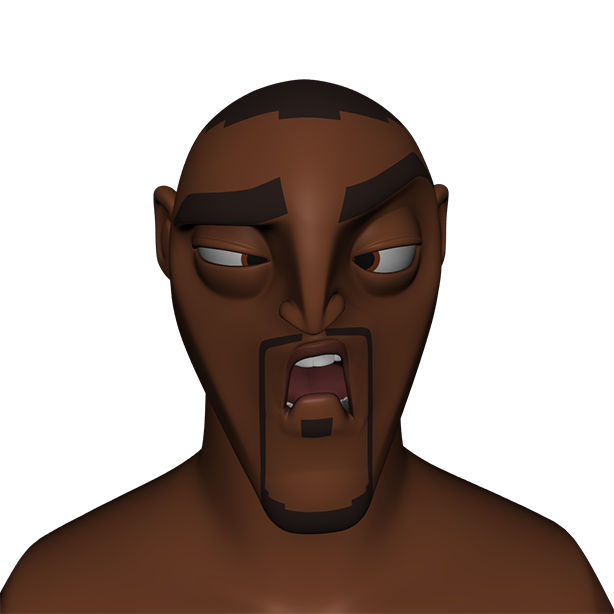} \\
\includegraphics[width=1\textwidth]{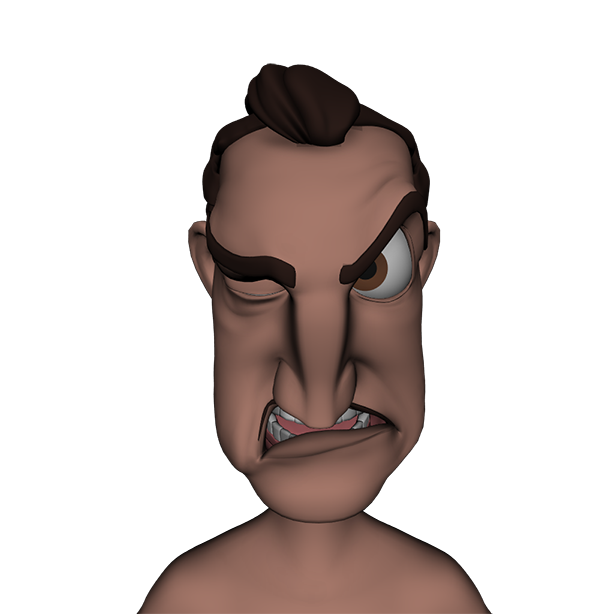} \\
\includegraphics[width=1\textwidth]{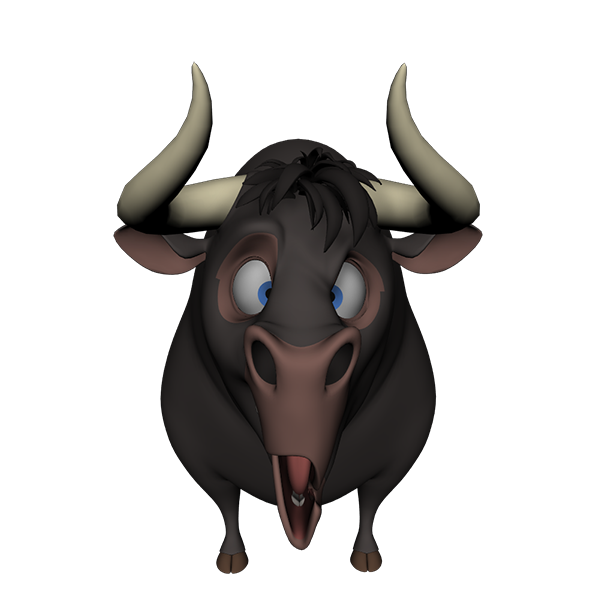}
\vspace{-2em}
\end{minipage}
}\hspace{2mm}
\subfigure{
\begin{minipage}[b]{0.1\textwidth}
\includegraphics[width=1\textwidth]{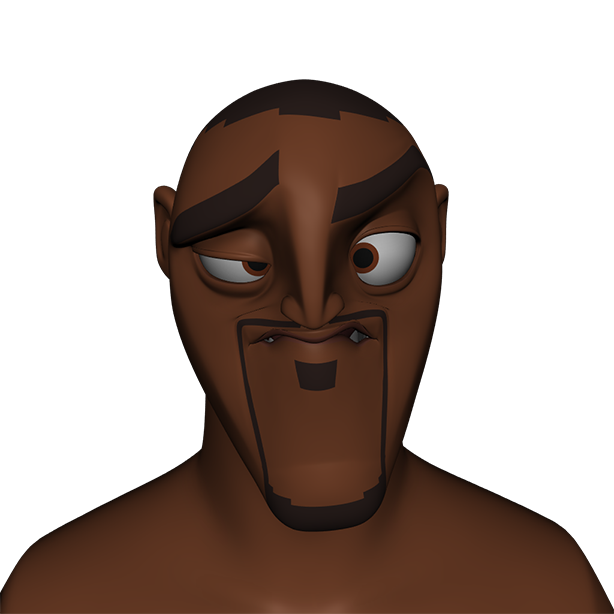} \\
\includegraphics[width=1\textwidth]{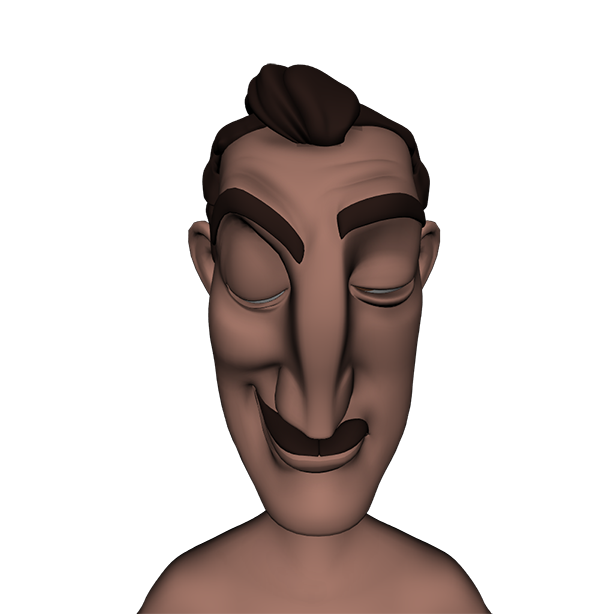} \\
\includegraphics[width=1\textwidth]{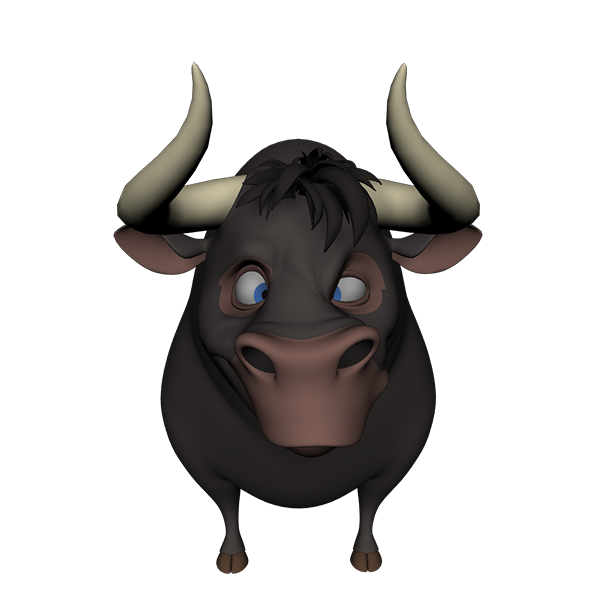}
\vspace{-2em}
\end{minipage}
}\hspace{2mm}
\subfigure{
\begin{minipage}[b]{0.1\textwidth}
\includegraphics[width=1\textwidth]{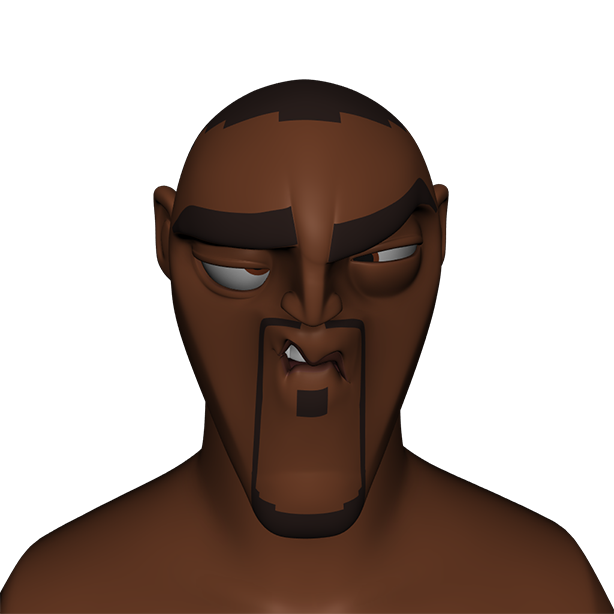} \\
\includegraphics[width=1\textwidth]{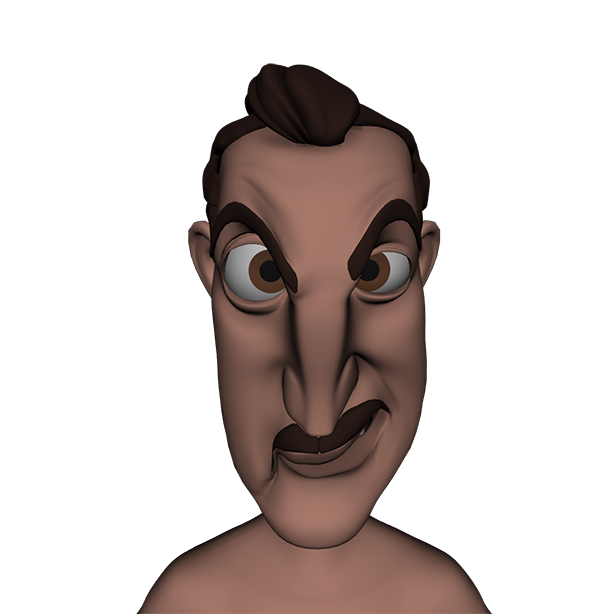} \\
\includegraphics[width=1\textwidth]{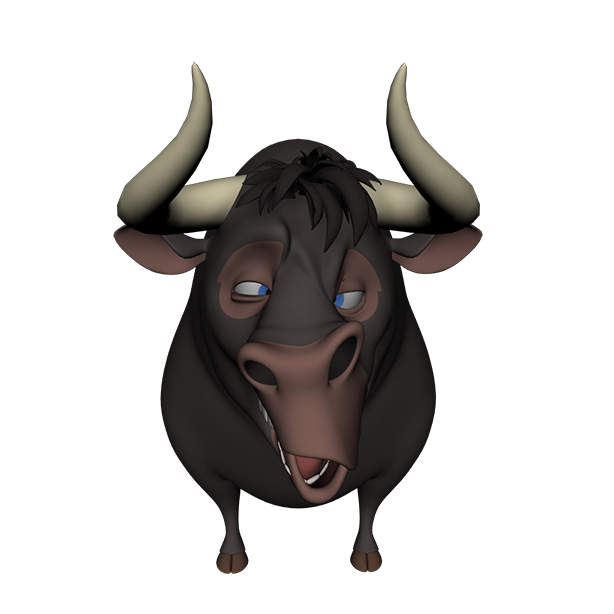}
\vspace{-2em}
\end{minipage}
}\hspace{2mm}
\subfigure{
\begin{minipage}[b]{0.1\textwidth}
\includegraphics[width=1\textwidth]{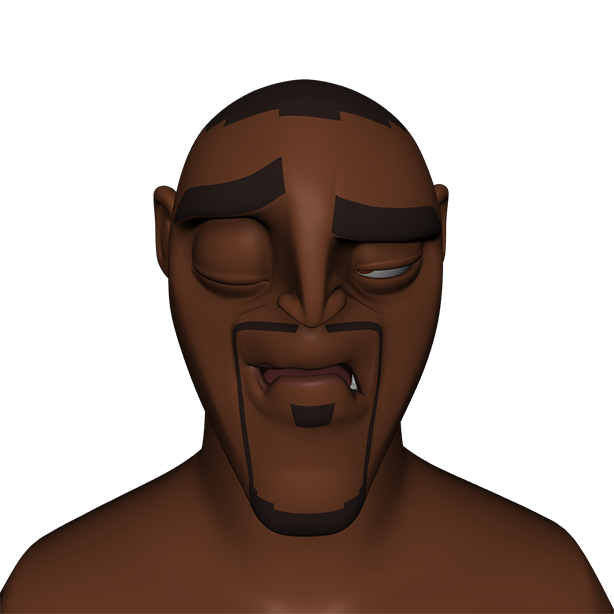} \\
\includegraphics[width=1\textwidth]{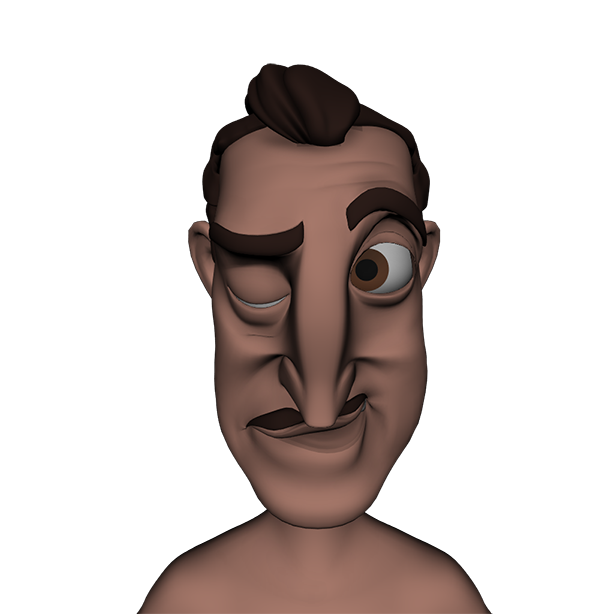} \\
\includegraphics[width=1\textwidth]{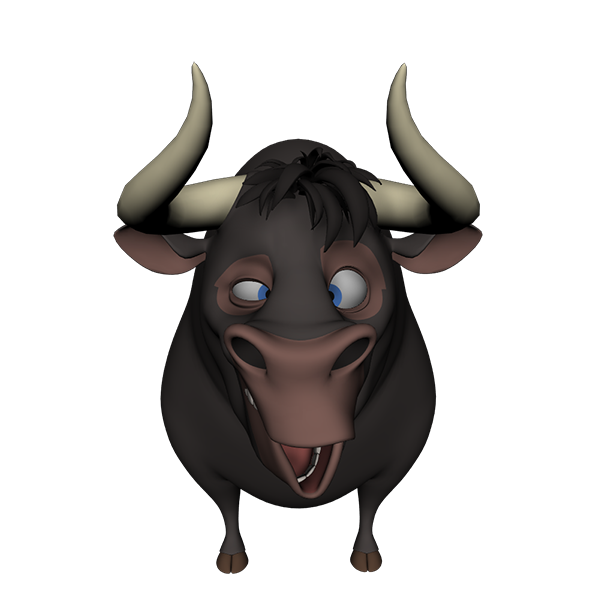}
\vspace{-2em}
\end{minipage}
}\hspace{2mm}
\subfigure{
\begin{minipage}[b]{0.1\textwidth}
\includegraphics[width=1\textwidth]{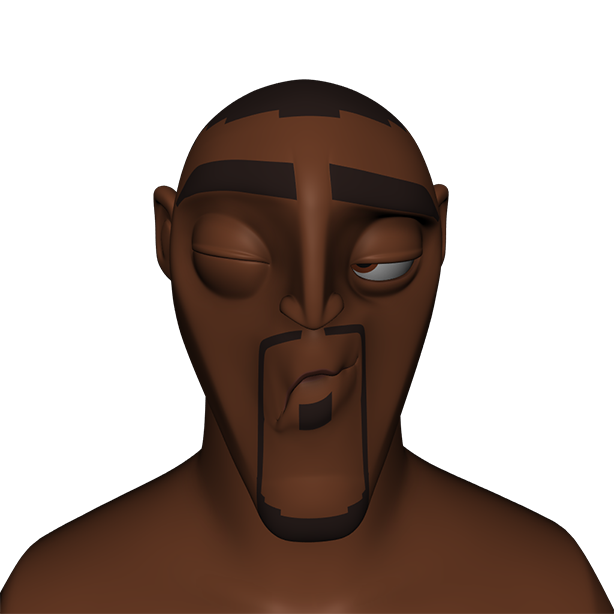} \\
\includegraphics[width=1\textwidth]{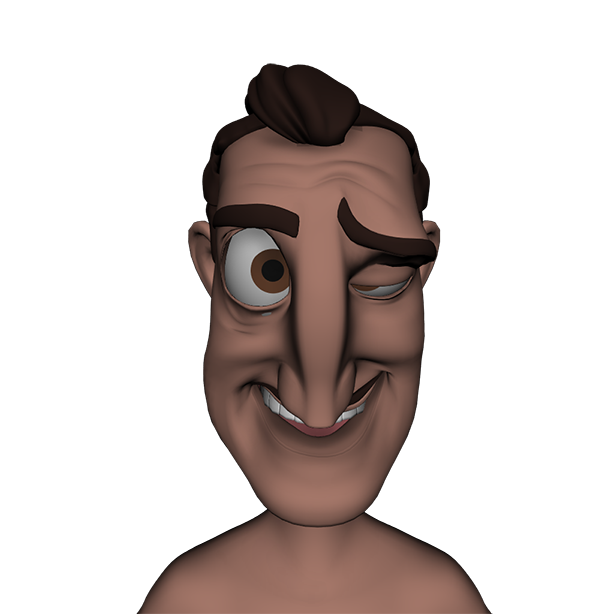} \\
\includegraphics[width=1\textwidth]{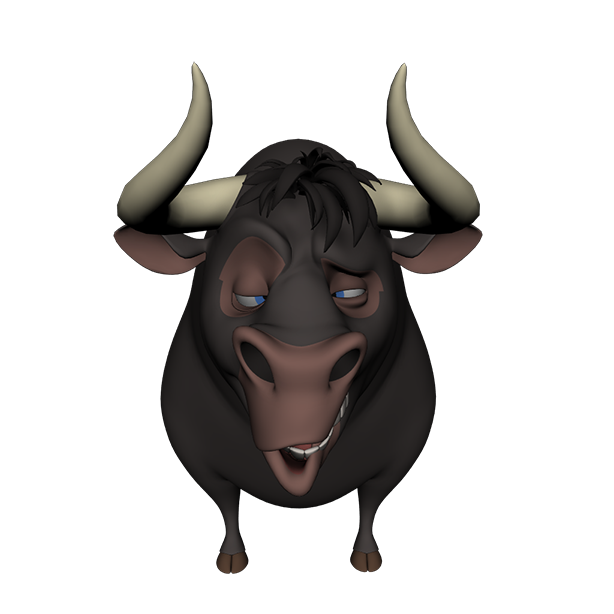}
\vspace{-2em}
\end{minipage}
}
\vspace{-1em}
\caption{\label{fig:exampleposes}Training poses for production characters \textit{Agent} (top), \textit{Matador} (middle) and \textit{Bull} (bottom). The poses are generated from a broad sampling of the rig parameter space. Although many look implausible, they are necessary to capture the full space accurately without assumptions on the artist's control range. }
\vspace{-1em}
\end{figure*}

We use three production face rigs for experiments and evaluation (see Table \ref{table:characterstats}). For each rig, we take a truncated normal sampling of the rig parameters to generate 10000 random poses: 9800 for training and 200 for testing (Fig. \ref{fig:exampleposes}). The test poses are separated from the training data to avoid bias. The rig parameters of the test poses are fed into the trained network to produce the reconstructed deformation (Fig. \ref{fig:pipeline}). To evaluate the training performance we use two metrics: the MSE of the prediction error and the reconstruction error ($cm$). The MSE of the prediction error measures the difference between the ground truth and network output, while the reconstruction error measures the per-vertex absolute distance between the surface reconstruction and ground truth deformation. We evaluate the mean and maximum reconstruction errors calculated from the vertices among all test poses. The maximum error is a critical value to consider as a large localized error will render the animation pose unacceptable, regardless of the MSE.

Because face rigs precisely control the eyelid, eyebrow, and mouth behavior, and because these are the primary cues for expression, having high accuracy here is paramount. A slight difference in eyelid position changes the relative position of the pupil, which can change the audience perception of the pose from ``scheming'' to ``sleepy'', while a similar change in the lip position can go from ``slight smile'' - with the teeth slightly exposed - to ``sneer'', making any method that could not accurately differentiate between these poses unacceptable.

\begin{table}[tb]
\caption{\label{table:characterstats}Statistics for the three test models. }
\vspace{-1em}
\centering
\scalebox{1}{
\begin{tabular}{cccc}
\hline
 &Agent&Bull&Matador\\
\hline
Vertices           & 4403 & 3669 & 3211\\

Face Height (cm)        & 25.12 & 84.28 & 26.03\\

Face Width (cm)        & 21.27 & 67.00 & 20.45\\

Numerical Controls & 67  &  131& 121\\
Joint Controls     & 20  & 20 & 20 \\
Anchor             & 87  & 73  & 64 \\
Differential PC    & 220 & 183 & 160\\

\hline
\end{tabular}
}
\end{table}

\subsection{Evaluation for Differential Training} 
\begin{figure}[tb]
  \centering
  \includegraphics[width=0.9\linewidth]{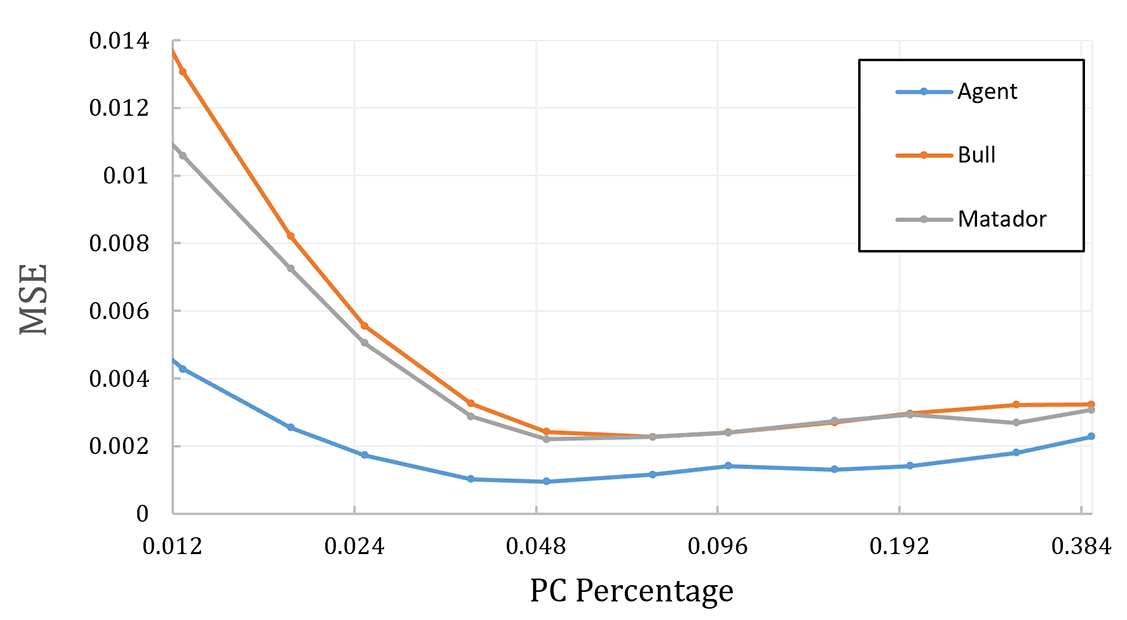}
  \vspace{-1em}
  \caption{\label{fig:pca} Prediction error of differential network with varying PC percentage}
  \vspace{-1em}
\end{figure}

We first evaluate how varying the number of principle components (PC) influences the differential training. We specify the PC number as a varying percentage of the mesh vertex count. Fig. \ref{fig:pca} shows the prediction error for three characters over 200 test poses. It is interesting to note that their MSE is minimized as the PC percentage approaches 5\%, regardless of the different number of vertices in their meshes. This suggests the optimal PC number is roughly proportional to the mesh vertex count. Further increasing the PC percentage does not lead to significant performance improvement, but instead  makes the network vulnerable to overfitting, shown by the slight increasing of the loss. Based on these observations, we set the PC number as 5\% of the mesh vertex count for differential training for the rest of our evaluation.

We use an ablation study to evaluate the influence of the hidden layer numbers, varying the number of fully connected layers from 1 to 5 while fixing the subspace network and anchor points. The prediction and reconstruction error for character \textit{Agent} for each of these is shown in Table \ref{table:ablation}. As shown, the prediction error decreases when the number of hidden layers increases, suggesting the improvement of network capacity for fitting. Also observable is the decrease of the reconstruction error,  but it is less significant compared with the reduction of prediction error,  suggesting that the accuracy of differential training is not the bottleneck for reconstruction.

\begin{table}[tb]
\caption{\label{table:ablation}Prediction error (differential) and reconstruction error (mean and maximum) of differential training with varying number of hidden layers and fixed subspace training. }
\vspace{-1em}
\centering
\scalebox{0.8}{
\begin{tabular}{cccccc}
\hline
 Layers & 1 & 2 & 3 & 4 & 5\\
\hline
Differential     & $2.58 \times 10^{-3}$  & $1.54 \times 10^{-3}$ & $1.10 \times 10^{-3}$  &  $1.03 \times 10^{-3}$ &  $9.47 \times 10^{-4}$ \\
Mean error       & 0.0240  & 0.0197 & 0.0189  &  0.0187 &  0.0182 \\
Max error        & 0.700   & 0.633  & 0.667   &  0.664  & 0.541\\
\hline
\end{tabular}
}
\end{table}

\subsection{Evaluation for Subspace Training} 
We use character \textit{Agent} to evaluate how the anchor points and subspace network influence the deformation approximation, considering different number of anchor points, selection methods and subspace network structures. For experiment purpose, we fix the differential training (4403 mesh vertices with 220PC) and only change the subspace network. We specify the number of anchor points as 1\%, 2\% and 5\% of the mesh vertex count, similar to our evaluation of PCA for differential training. We report both the prediction and reconstruction error in Table \ref{table:anchorcmp}. Notice we increase the percentage by adding new anchor points into the existing ones instead of selecting a new group. To compare the network structure, we conduct the subspace training using a single network instead of the subspace mini-networks (``2\%Single''). The single network takes the entire vectorized features as input and outputs the deformation of all anchor points together. To compare different anchor point selection methods, we use a new group of anchor points around the scalp with less significant deformation (``2\% Scalp''). Notice the original group of anchor points are selected on the face to cover major facial features with large deformation, as discussed in Section 3.2. 

As observed, increasing the number of anchor points leads to higher prediction error since the network performs better fitting when the dimension is low. However, the reconstruction error stays roughly the same when the number of anchor point gets larger, because increasing the number of anchor points can improve the Laplacian matrix condition for reconstruction, which balances the increase of prediction error. We use 2\% anchor points as a middle point for our implementation and the rest of the evaluation. 

For the network comparison, both the prediction and reconstruction error of the subspace mini-networks (``2\%'') are lower than the single network (``2\%Single''). We believe the dimension reduction is the reason for resulting performance improvement. The subspace mini-networks fit anchor points separately because they are disconnected and do not have direct spatial relationship, which enables better approximation. The single network, on the contrary, tries to learn the deformation of anchor points all at once, increasing the difficulty of fitting.

\begin{figure}
  \centering
  \includegraphics[width=0.95\linewidth]{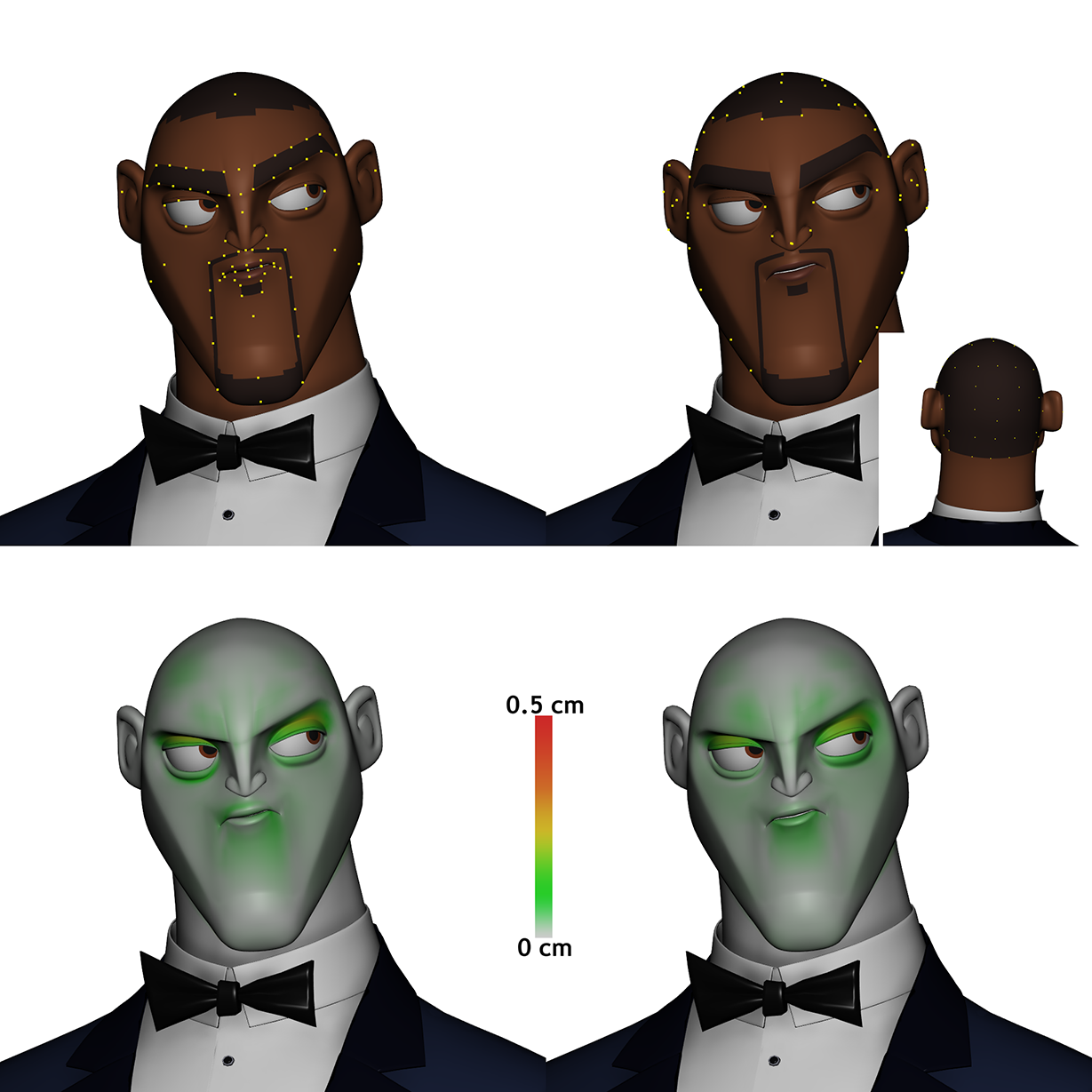}
  \vspace{-1em}
  \caption{\label{fig:anchorselection} Comparison of our method using anchors (in yellow) selected from the major facial features (left) vs. the less deformed scalp (right). }
\end{figure}

For different anchor point selection, we find using vertices with less deformation can cause larger reconstruction error even when the prediction error is smaller. The network has better performance because no deformation needs to be learned for those vertices, but they are not ideal for the reconstruction. Fig. \ref{fig:anchorselection} shows an example. As we can see, the deformation on mouth and eyelids are shifted when vertices on the scalp are selected as anchor points. Ideally, we want the anchor points to ``nail'' the deformed mesh in place and prevent large shifts or rotations for important face regions. Therefore, we select anchor points to cover major facial features with large deformation.

\begin{table}[tb]
\caption{\label{table:anchorcmp}Prediction error (subspace) and reconstruction error (mean and max) of the subspace training with varying anchor percentage with fixed differential training.}
\vspace{-1em}
\centering
\scalebox{0.7}{
\begin{tabular}{c c >{\bfseries}c c c c}
\hline
 & 1\% & 2\% & 2\%(Single) & 2\%(Scalp) & 5\% \\
\hline
Subspace    & $1.35 \times 10^{-3}$ & \boldmath $1.71 \times 10^{-3}$\unboldmath  &  $7.42 \times 10^{-3}$ & $9.37 \times 10^{-4}$ & $1.73 \times 10^{-3}$\\
Mean error  &  0.0207   & 0.0186  &  0.0336 & 0.0192 & 0.0158 \\
Max error   &  0.524    & 0.517   &  0.657  & 0.562  &  0.577 \\
\hline
\end{tabular}
}
\end{table}

\subsection{Results}
\begin{figure}
\centering
\includegraphics[width=1.0\linewidth]{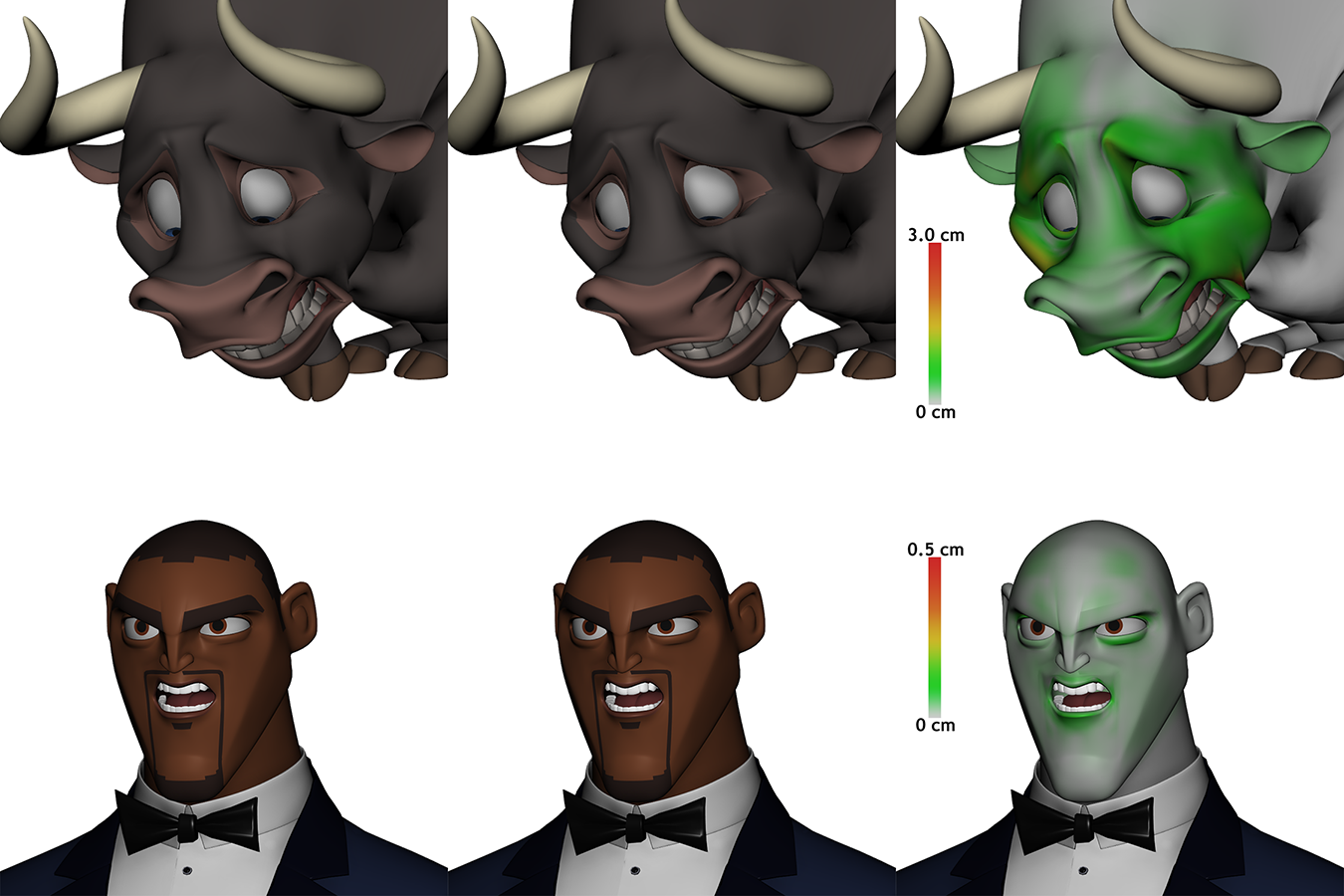} \\
\vspace{-1em}
\caption{\label{fig:result} Side-by-side comparison of ground truth (left), our approximation (center), and heatmap indicating per-vertex distance error in cm (right).}
\end{figure}

In this section, we evaluate the accuracy of deformation reconstruction using well-animated poses from production. We evaluate the mean and max reconstruction errors over a series well-animated production sequences, where the deformations are much more exaggerated and dynamic. We present the quantitative results in Table \ref{table:errors}. The deformations of character \textit{bull} are observed with larger errors because we test it on the most extreme animation sequence. Fig. \ref{fig:result} shows an example for the character and please refer to the supplemental video for detailed comparison. In general, our method can accurately reconstruct mesh surface with mean errors smaller than 0.6\% and max error smaller than 6\% of the size of the character faces.

As a data-driven solution, the accuracy of our model largely relies on sufficient training data. To evaluate how the training size influence the performance, we alternatively reduce the size for character \textit{Agent} to be 25\%, 50\% and 75\% of the original dataset while keeping the test data unchanged (200 randomly generated poses). We present both the prediction errors for the differential and subspace training and reconstruction errors in Table \ref{table:size}. Indeed, the increasing of training data will boost the performance. However, the improvement is not very significant when increasing the size over 75\%. 

\begin{table}[tb]
\caption{\label{table:size}Prediction errors (Differential and Subspace) and reconstruction errors (Mean and Max) for the tests with different training data size.}
\vspace{-1em}
\centering
\scalebox{0.9}{
\begin{tabular}{ccccc}
\hline
                           &  25\%   &  50\%  &75\%     & 100\%\\
\hline
Differential   & $2.66\times 10^{-3}$ & $1.57\times 10^{-3}$ & $1.03\times 10^{-3}$ & $1.53\times 10^{-3}$ \\
Subspace       & $3.91\times 10^{-3}$ & $2.81\times 10^{-3}$ & $2.07\times 10^{-3}$ & $1.71\times 10^{-3}$ \\
Mean error                 &0.0301 & 0.0246 & 0.0195 & 0.0186\\
Max error                  & 0.891  & 0.819  & 0.740 & 0.517\\
\hline
\end{tabular}
}
\end{table}

\begin{table}[tb]
\caption{\label{table:errors}Mean and max reconstruction absolute errors evaluated on the well-animated production sequences, and as a percentage of face height.}
\vspace{-1em}
\centering
\scalebox{1}{
\begin{tabular}{cccc}
\hline
 &Agent&Bull&Matador\\
\hline
Mean error         & 0.032 & 0.512 & 0.087\\
Percentage         & 0.127\% & 0.607\% & 0.334\% \\ 
Max error          & 0.630  & 4.682  & 0.782\\
Percentage         & 2.50\% & 5.55\% & 3.00\% \\ 
Number of Poses    & 808     & 249     & 359 \\
\hline
\end{tabular}
}
\end{table}

\subsection{Comparison} 

We first compare the accuracy of facial deformation approximation with previous methods. Then we apply our method to body rigs and compare the results with Bailey et al. \shortcite{bailey2018fast}.

\subsubsection{Facial Deformation Comparison} We compare our method with linear blend skinning (LBS), PCA with linear regression (PCA), local Cartesian coordinates training using our model (Local) and Meyer et al. \shortcite{meyer2007key} (KPSA). KPSA is an example-based deformation approximation method, which uses the deformation of key points as input to PCA to derive vertex positions for the entire mesh. The quality of the training data significantly influences the accuracy of the deformation, and their method relies on evaluating the original deformer stack to determine the key points on the fly. For the Local model,  we apply the same differential network with PCA directly on the vertex local offsets without converting them into differential coordinates. No subspace learning and reconstruction is required for this model. We use it to compare the differential training and evaluate the contribution of mesh representation. We use the same set of randomly-generated training poses as used by our model to train both KPSA and the Local model, and we perform evaluation on the same well-animated sequences introduced in the last subsection.

\begin{figure*}[tb]
\centering
\subfigure{
\begin{minipage}[b]{0.16\textwidth}
\includegraphics[width=1\textwidth]{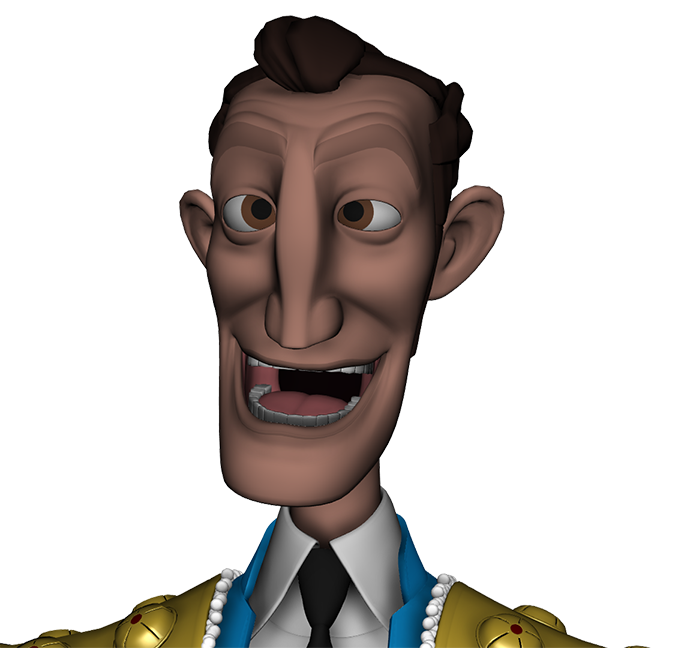}\\\
\includegraphics[width=1\textwidth]{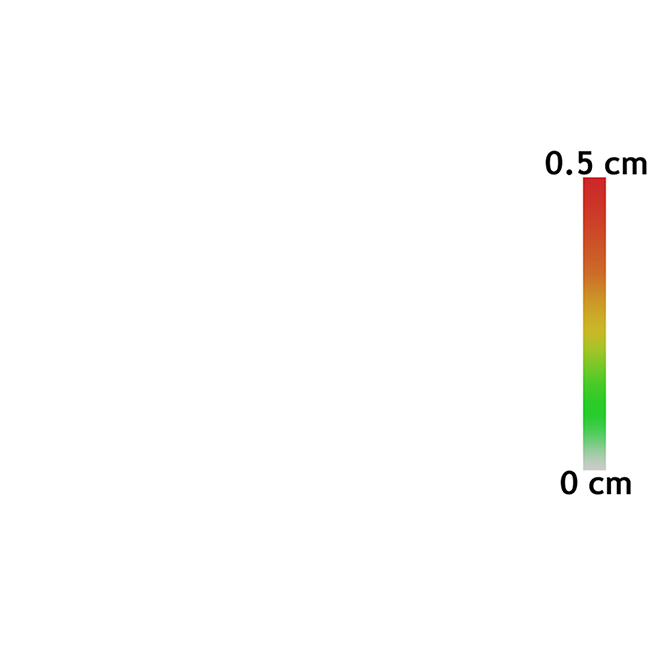}
\vspace{-2em}
\caption*{Ground Truth}
\end{minipage}
}\hspace{-2mm}
\subfigure{
\begin{minipage}[b]{0.16\textwidth}
\includegraphics[width=1\textwidth]{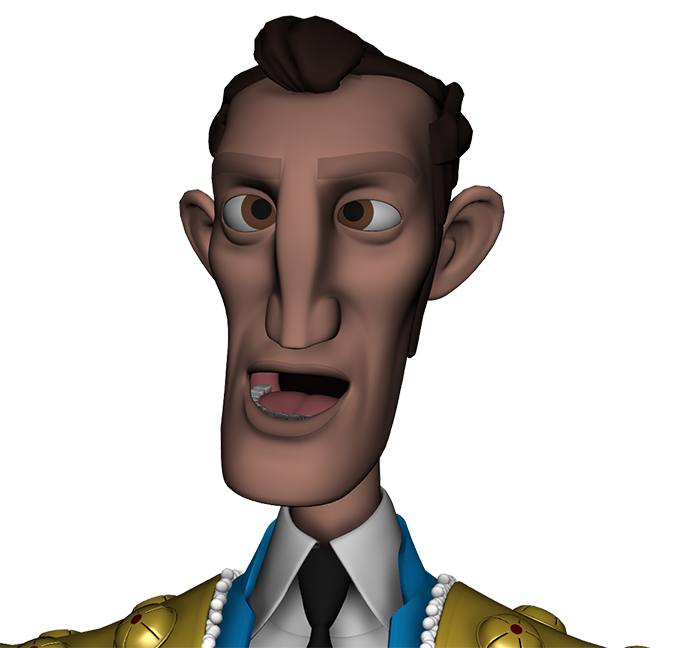}\\
\includegraphics[width=1\textwidth]{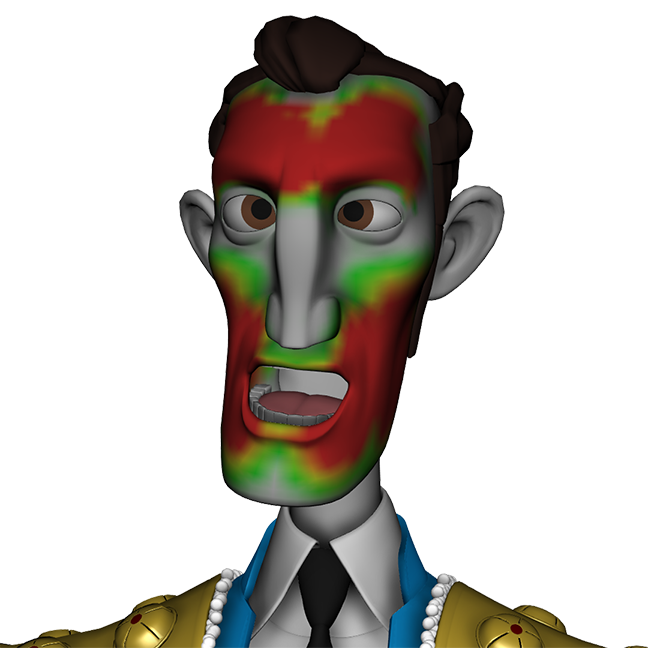}
\vspace{-2em}
\caption*{LBS}
\end{minipage}
}\hspace{-2mm}
\subfigure{
\begin{minipage}[b]{0.16\textwidth}
\includegraphics[width=1\textwidth]{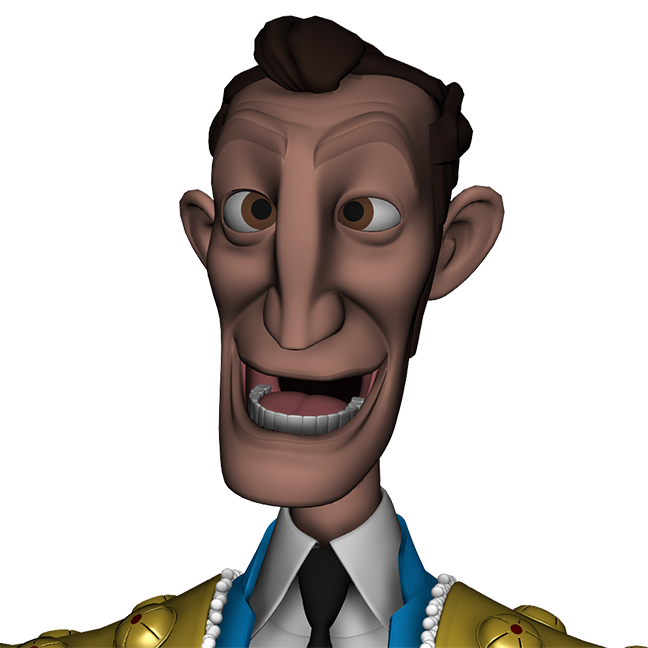} \\
\includegraphics[width=1\textwidth]{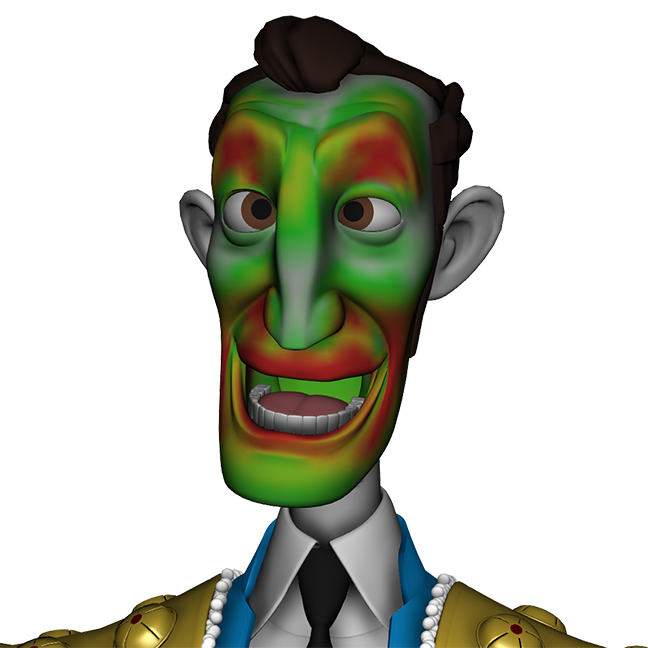}
\vspace{-2em}
\caption*{PCA}
\end{minipage}
}\hspace{-2mm}
\subfigure{
\begin{minipage}[b]{0.16\textwidth}
\includegraphics[width=1\textwidth]{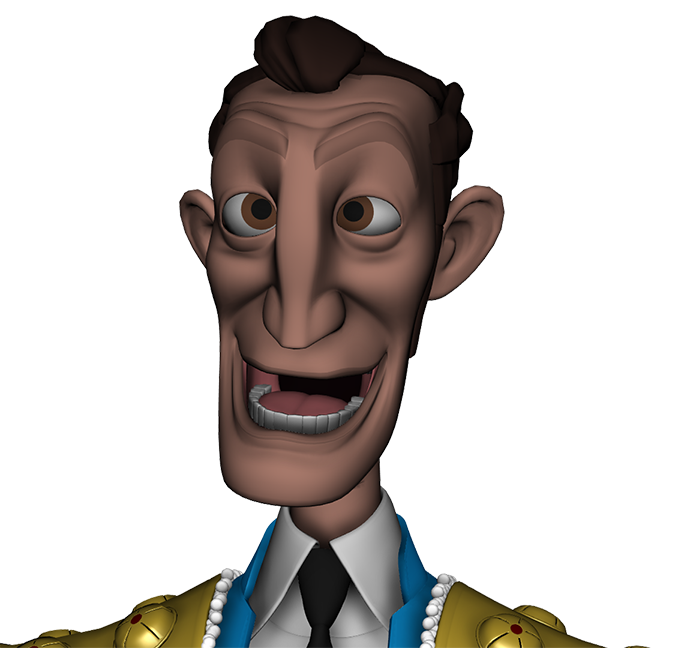} \\
\includegraphics[width=1\textwidth]{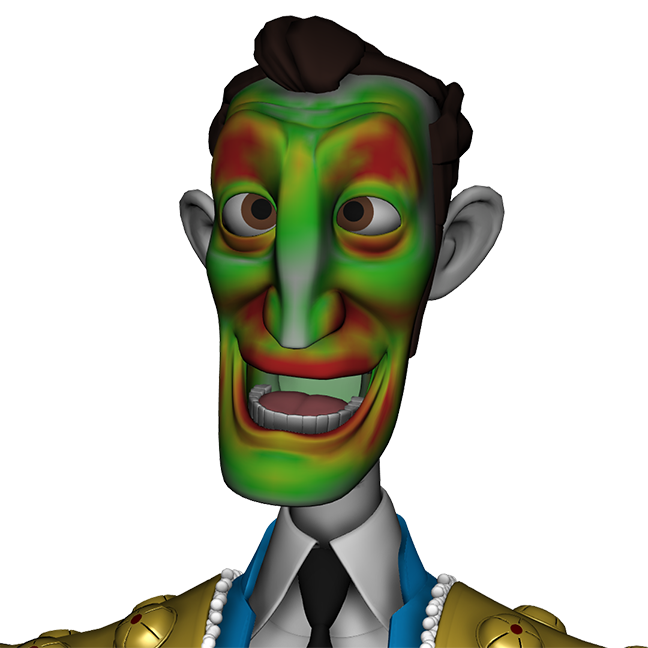}
\vspace{-2em}
\caption*{Local}
\end{minipage}
}\hspace{-2mm}
\subfigure{
\begin{minipage}[b]{0.16\textwidth}
\includegraphics[width=1\textwidth]{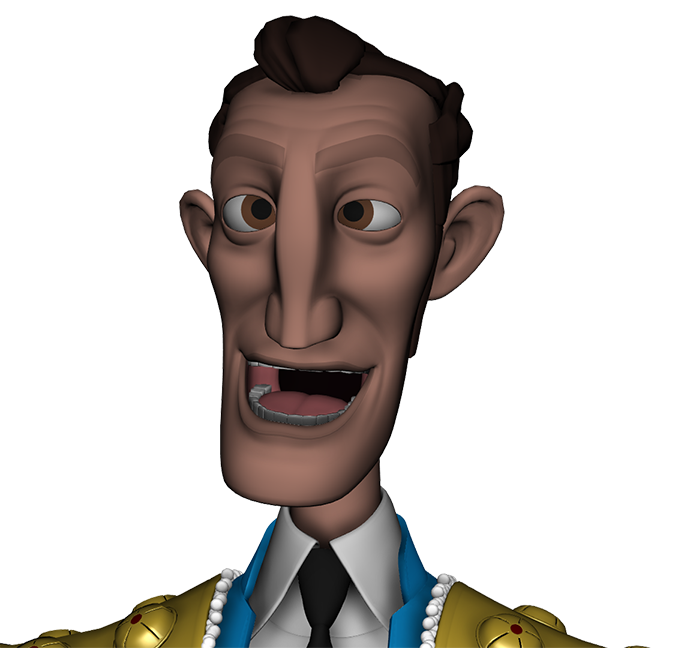}\\
\includegraphics[width=1\textwidth]{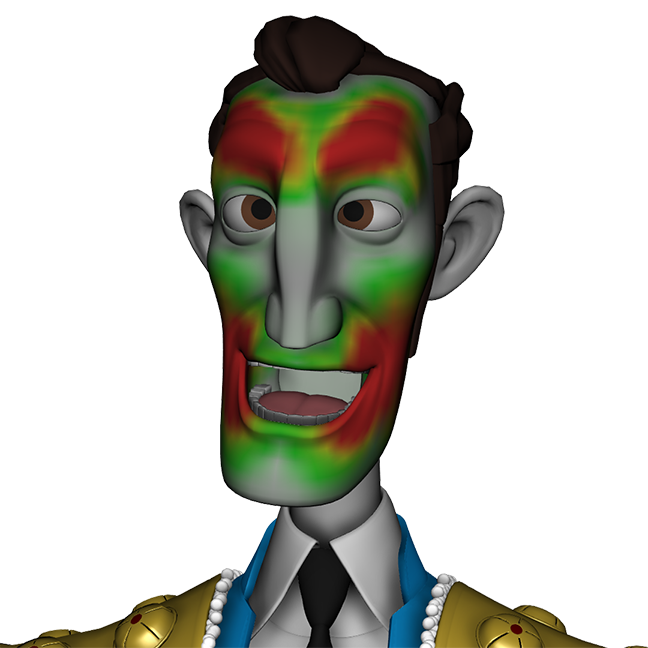}
\vspace{-2em}
\caption*{KPSA}
\end{minipage}
}\hspace{-2mm}
\subfigure{
\begin{minipage}[b]{0.16\textwidth}
\includegraphics[width=1\textwidth]{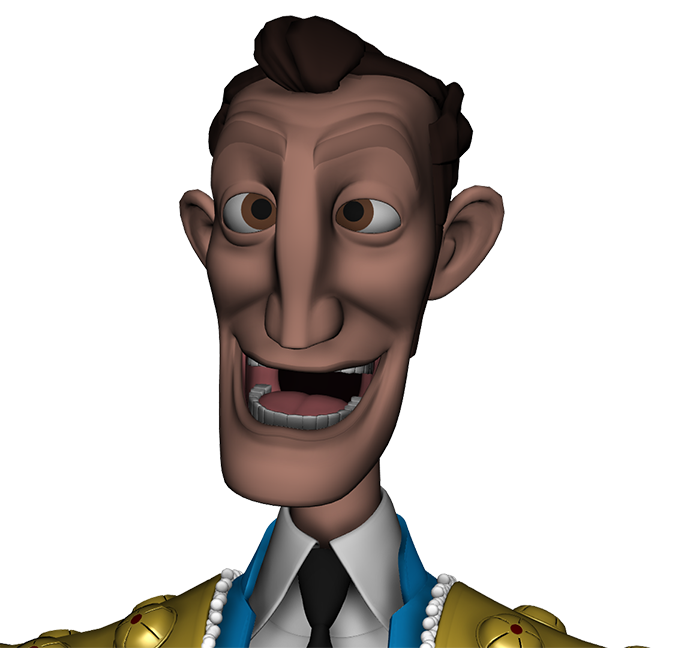}\\
\includegraphics[width=1\textwidth]{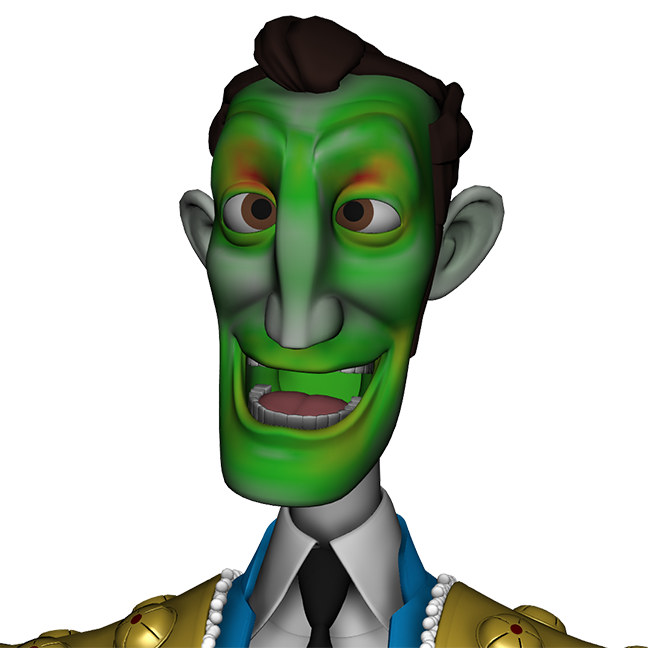}
\vspace{-2em}
\caption*{Ours}
\end{minipage}
}
\vspace{-1em}
\caption{\label{fig:facecmp} Comparisons for facial deformation between ground truth, Linear Blend Skinning (LBS), PCA with linear regression, local offset training, KPSA, and our method using a well-animated pose from production.}
\vspace{-1em}
\end{figure*}

We report the reconstruction error in Table \ref{tab:facecmp} and provide visual comparison in Fig. \ref{fig:facecmp}. As observed, our method outperforms the other four methods in both quantitative and visualized results. We use the result of LBS as a base-line as it does not provide any nonlinear deformation. From the heat map, we can see that the Local model fails to capture the local deformation on the eyelids and the mouth is shifted. This is because no neighbor vertex information is embedded in the local offset, which makes it difficult for the network to predict the local deformation. For KPSA, it fails to reconstruct the deformation in the eyebrow region and the corner of the lips, even though with a substantial increase in the number of key points (274) and basis vectors (200) used in the original example. The relatively poor performance is caused by the linear reconstruction of training data, which could only provide a limited range and a fixed dimension for the approximated deformation. Once the target pose is out of the dimension defined by the PCA, it is difficult for that method to achieve high reconstruction accuracy. Additionally, the key points still need to be driven by the original rig. In comparison, our method can accurately capture the local deformation because of the error characteristics of the differential coordinates. Due to the nonlinear fitting capability of deep neural networks, our method can use randomly generated data for training and approximate deformation with a much larger range.

\begin{table}[t]
\caption{Mean and max reconstruction errors using our method compared with Linear Blend Skinning (LBS), PCA with linear regression, our model using local offset for training (Local) and Meyer et al. \shortcite{meyer2007key} (KPSA). The comparison is shown for a set of test poses from a well-animated production sequence.}
\centering
\scalebox{1.1}{
\begin{tabular}{c|c|c|c|c|c|c|}
\cline{2-7}
\multirow{2}{*}{} & \multicolumn{2}{c|}{Agent} & \multicolumn{2}{c|}{Bull} & \multicolumn{2}{c|}{Matador} \\ \cline{2-7}
        & \multicolumn{1}{c}{Mean}  & Max   & \multicolumn{1}{c}{Mean}  & Max    & \multicolumn{1}{c}{Mean}   & Max \\ \cline{2-7}
LBS    & 0.174 & 3.228 & 1.672 & 23.56  & 0.228 & 4.261 \\ \cline{2-7}
PCA    & 0.073 & 1.980 & 0.848 & 8.367  & 0.158 & 1.533 \\ \cline{2-7}
Local  & 0.072 & 0.689 & 0.521 & 5.779  & 0.155  & 1.106  \\ \cline{2-7}
KPSA   & 0.061 & 1.623 & 2.115 & 34.25  & 0.089 & 1.664  \\\cline{2-7}
Ours   & 0.032 & 0.630 & 0.512 & 4.682  & 0.087 & 0.782 \\\cline{2-7}
\end{tabular}
}
\vspace{.75em}
\label{tab:facecmp}
\vspace{-2em}
\end{table}

\subsubsection{Body Deformation Comparison} We demonstrate our method applied to body deformation approximation and compare our results with Bailey et al. \shortcite{bailey2018fast}. We use character \textit{Agent} as the example for comparison. The character's height is 200.25 cm. The body contains 4908 vertices and the rig includes 107 joints controls with hand joints excluded. We use the same training method and network structures mentioned in Section 3.2 and generate random poses for body rig as training data. Since the body rig does not include numerical controls, we remove them from input and only vectorize the joint controls. We use 245 PCs for the differential training and select 118 anchor points that are well-distributed around all the joints of the body. For Bailey et al. \shortcite{bailey2018fast}, denoted as FDDA, we follow their methods to train multiple small networks (2 hidden layers with 128 units), each of which corresponds to a joint control and predicts the nonlinear deformation of the neighbor vertices in local coordinates. We generate 9800 random poses using the method described in Section 3.3 as training data and perform the evaluation using 189 poses from a well-animated production sequence for all three models.

\begin{wraptable}{r}{4cm}
\vspace{-10pt}
\hspace{-10pt}
\begin{tabular}{r|cc}
              & Mean  & Max \\ \hline
Ours          & 0.217 & 4.17   \\
FDDA          & 0.263 & 6.41    \\
\end{tabular}
\hspace{-10pt}
\vspace{-10pt}
\end{wraptable} 

\begin{figure*}[tb]
\centering
\vspace{-2em}
\subfigure{
\begin{minipage}[b]{0.27\textwidth}
\includegraphics[width=1\textwidth]{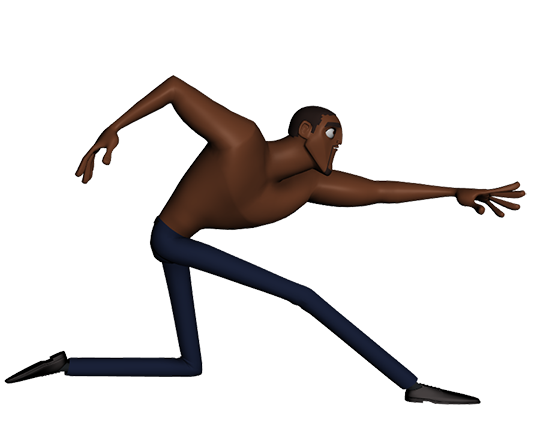}\\
\includegraphics[width=1\textwidth]{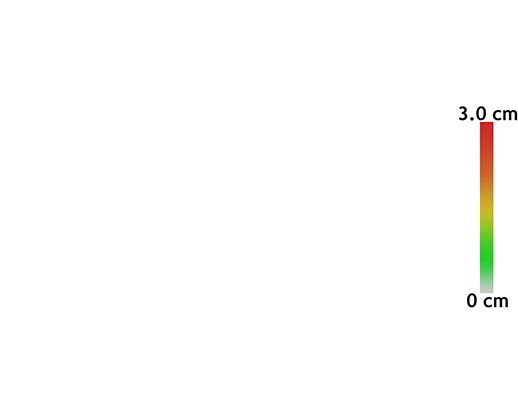}
\end{minipage}
}\hspace{-5mm}
\subfigure{
\begin{minipage}[b]{0.27\textwidth}
\includegraphics[width=1\textwidth]{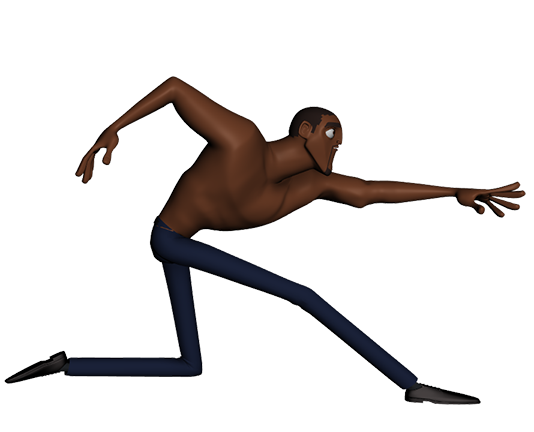} \\
\includegraphics[width=1\textwidth]{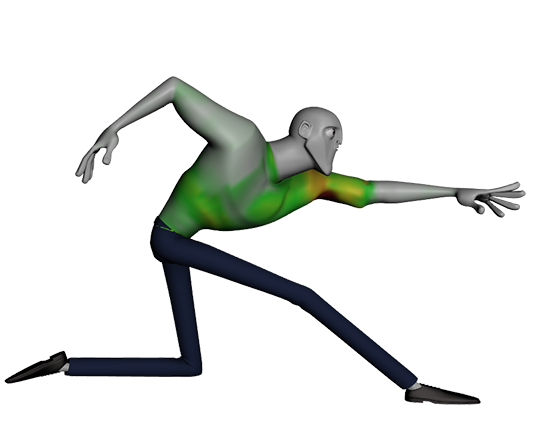}
\end{minipage}
}\hspace{-5mm}
\subfigure{
\begin{minipage}[b]{0.27\textwidth}
\includegraphics[width=1\textwidth]{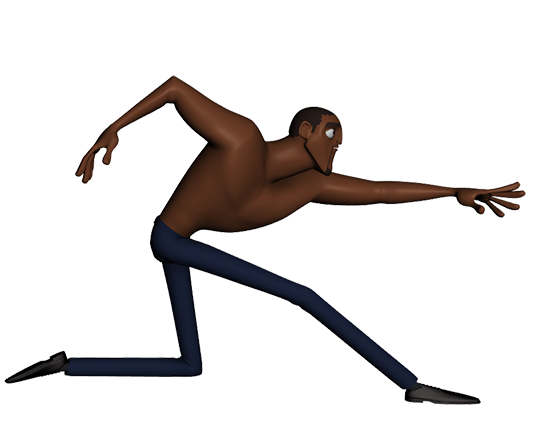}\\
\includegraphics[width=1\textwidth]{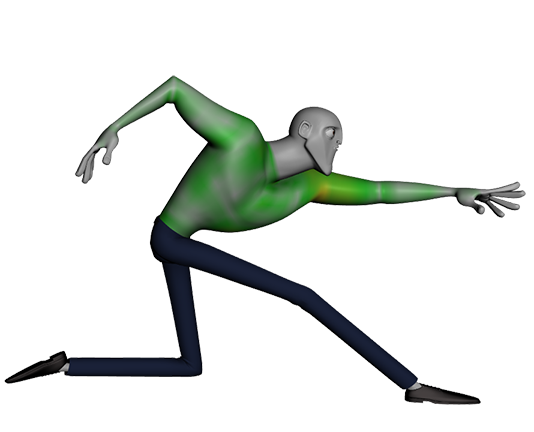}
\end{minipage}
}\hspace{-2mm}
\subfigure{
\begin{minipage}[b]{0.27\textwidth}
\includegraphics[width=1\textwidth]{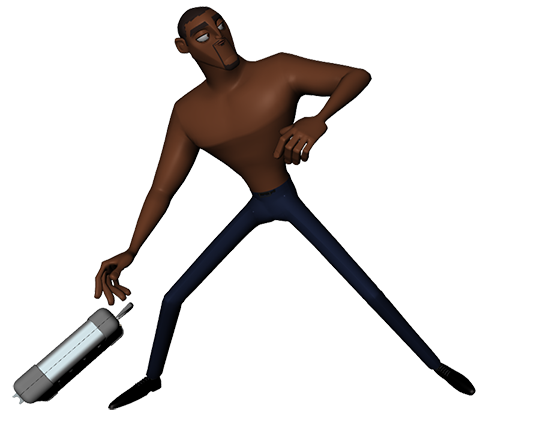}\\
\includegraphics[width=1\textwidth]{image/white_body_cmp_withscale.png}
\vspace{-2em}
\caption*{Ground Truth}
\end{minipage}
}\hspace{-5mm}
\subfigure{
\begin{minipage}[b]{0.27\textwidth}
\includegraphics[width=1\textwidth]{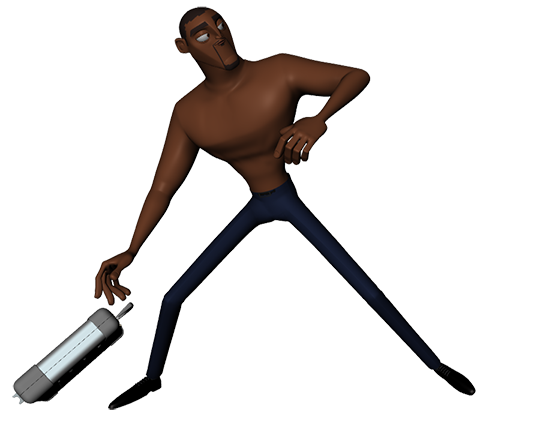} \\
\includegraphics[width=1\textwidth]{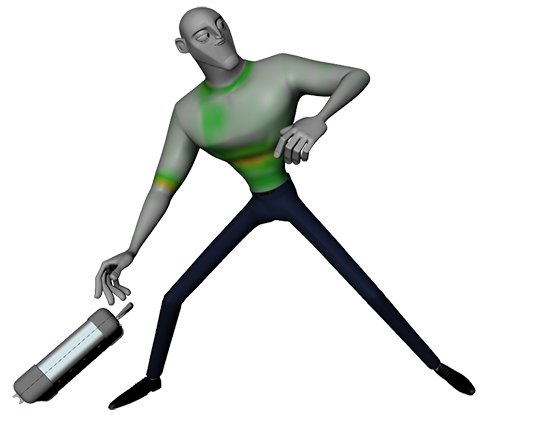}
\vspace{-2em}
\caption*{FDDA}
\end{minipage}
}\hspace{-5mm}
\subfigure{
\begin{minipage}[b]{0.27\textwidth}
\includegraphics[width=1\textwidth]{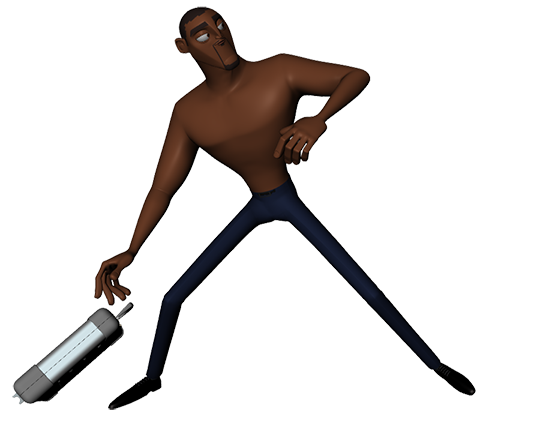}\\
\includegraphics[width=1\textwidth]{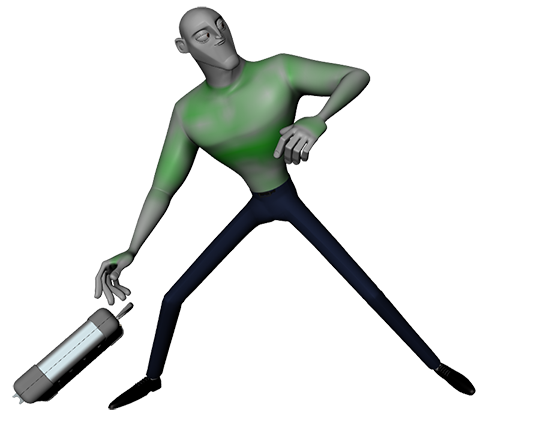}
\vspace{-2em}
\caption*{Ours}
\end{minipage}
}
  \vspace{-1em}
  \caption{\label{fig:bodycmp}  Comparisons for body deformation between the ground truth, Bailey et al. \shortcite{bailey2018fast} (FDDA) and our method using well-animated poses.}
\end{figure*}

We report the mean and max reconstruction error in the inline table and we show deformation results in Fig \ref{fig:bodycmp}. The results indicate that our method outperforms the FDDA method, especially for the maximum error. Using multiple networks for deformation approximation, FDDA suffers from discontinuity problem on torso and left arm. We can observe high errors on the connecting parts of the body since the vertices from the two parts are predicted by different networks. The discontinuity is caused by the slight change of joint scales in the evaluation sequence, which does not show up in the training data. Due to the local joint input and small-scale network, FDDA suffers from overfitting to the training data and is sensitive to new values. Our method uses a deeper network with a much larger input size, which increases the capacity and makes the network less sensitive to the unseen scaling change of a couple joints. Since our method also uses small networks for subspace training, there might be some anchors that are affected by the scaling. But due to the least square reconstruction, the local error is nicely distributed as low frequency error and is much less noticeable. Increasing the network size for FDDA may improve the overall performance, however evaluating a large number of deeper networks (40 in our case) would cause significant performance downgrade. 

\begin{figure}[tb]
\centering
\hspace{-3mm}
\subfigure{
\begin{minipage}[b]{0.25\textwidth}
\includegraphics[width=1\textwidth]{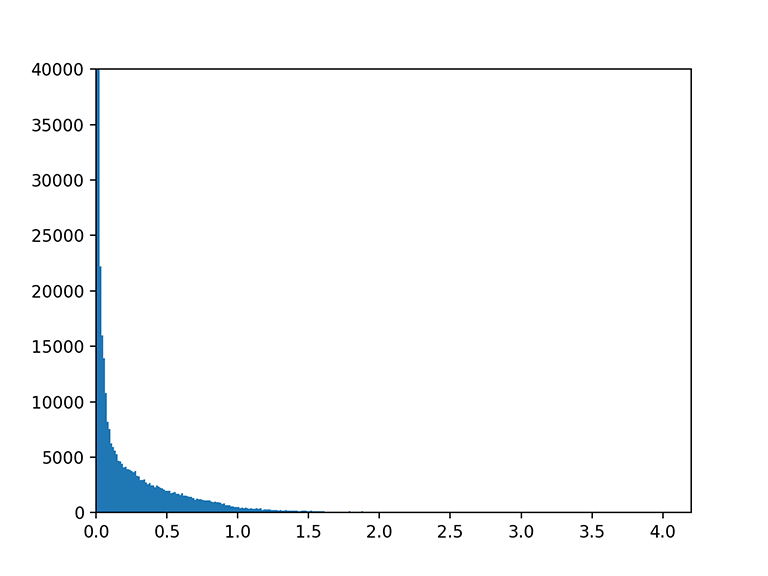}
\vspace{-3em}
\caption*{FDDA}
\end{minipage}
}\hspace{-6mm}
\subfigure{
\begin{minipage}[b]{0.25\textwidth}
\includegraphics[width=1\textwidth]{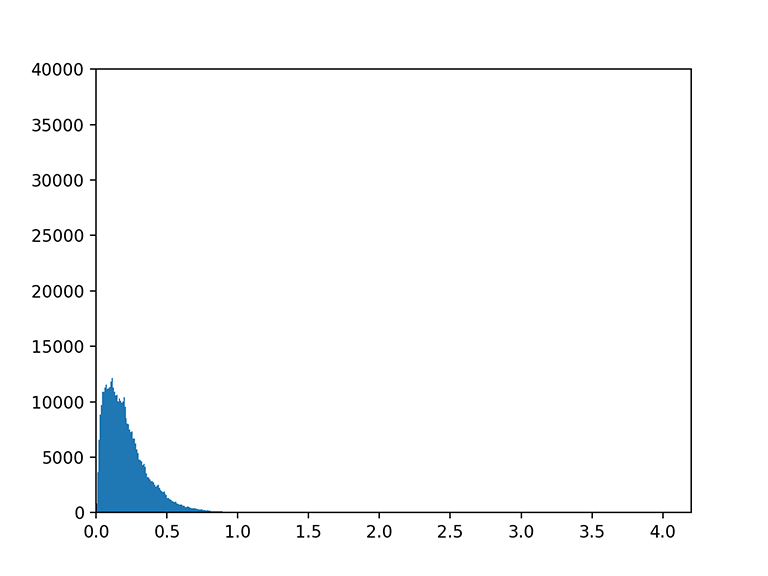}
\vspace{-3em}
\caption*{Ours}
\end{minipage}
}
  \vspace{-2.3em}
  \caption{\label{fig:histlancebody} Comparison of error distribution for body deformation using well-animated pose from production.}
  \vspace{.5em}
\end{figure}

Fig. \ref{fig:histlancebody} shows the error distribution of each model. As observed, the error distribution of our model is compressed to the lower range while the distribution of FDDA extends to large errors. Although the two methods have similar mean error, this observation suggests that our method can provide smooth approximation results with smaller maximum errors, and avoids inappropriate deformation.

\section{Conclusion}

In this paper we have presented a learning-based solution to capture facial deformation for rigs with high accuracy. Our method uses differential coordinates and a learned subspace to reconstruct smooth nonlinear facial deformation. We have demonstrated the robustness of our method on a wide range of animated poses. Our method compares favorably with existing solutions for both facial and body deformation. We also have successfully integrated our solution into the production pipeline.

Our work has limitations that we wish to investigate in the future. First, our method needs manually-selected anchor points for subspace training and reconstruction. It would be interesting to investigate methods for inferring the anchor points based on the characteristics of the facial mesh and training poses. Second, as a deep learning based approach, a model must to be trained for every character with different rig behavior or mesh topology. We would like to explore the possibility of integrating a high level super-rig into our method to provide a single model for different characters.

\bibliographystyle{ACM-Reference-Format}
\bibliography{main-bibliography}


\begin{thebibliography}{52}


\ifx \showCODEN    \undefined \def \showCODEN     #1{\unskip}     \fi
\ifx \showDOI      \undefined \def \showDOI       #1{#1}\fi
\ifx \showISBNx    \undefined \def \showISBNx     #1{\unskip}     \fi
\ifx \showISBNxiii \undefined \def \showISBNxiii  #1{\unskip}     \fi
\ifx \showISSN     \undefined \def \showISSN      #1{\unskip}     \fi
\ifx \showLCCN     \undefined \def \showLCCN      #1{\unskip}     \fi
\ifx \shownote     \undefined \def \shownote      #1{#1}          \fi
\ifx \showarticletitle \undefined \def \showarticletitle #1{#1}   \fi
\ifx \showURL      \undefined \def \showURL       {\relax}        \fi
\providecommand\bibfield[2]{#2}
\providecommand\bibinfo[2]{#2}
\providecommand\natexlab[1]{#1}
\providecommand\showeprint[2][]{arXiv:#2}

\bibitem[\protect\citeauthoryear{An, Kim, and James}{An et~al\mbox{.}}{2008}]%
        {an2008optimizing}
\bibfield{author}{\bibinfo{person}{Steven~S An}, \bibinfo{person}{Theodore
  Kim}, {and} \bibinfo{person}{Doug~L James}.} \bibinfo{year}{2008}\natexlab{}.
\newblock \showarticletitle{Optimizing cubature for efficient integration of
  subspace deformations}. In \bibinfo{booktitle}{\emph{ACM transactions on
  graphics (TOG)}}, Vol.~\bibinfo{volume}{27}. ACM, \bibinfo{pages}{165}.
\newblock


\bibitem[\protect\citeauthoryear{Bailey, Otte, Dilorenzo, and O'Brien}{Bailey
  et~al\mbox{.}}{2018}]%
        {bailey2018fast}
\bibfield{author}{\bibinfo{person}{Stephen~W Bailey}, \bibinfo{person}{Dave
  Otte}, \bibinfo{person}{Paul Dilorenzo}, {and} \bibinfo{person}{James~F
  O'Brien}.} \bibinfo{year}{2018}\natexlab{}.
\newblock \showarticletitle{Fast and deep deformation approximations}.
\newblock \bibinfo{journal}{\emph{ACM Transactions on Graphics (TOG)}}
  \bibinfo{volume}{37}, \bibinfo{number}{4} (\bibinfo{year}{2018}),
  \bibinfo{pages}{119}.
\newblock


\bibitem[\protect\citeauthoryear{Barbi{\v{c}} and James}{Barbi{\v{c}} and
  James}{2005}]%
        {barbivc2005real}
\bibfield{author}{\bibinfo{person}{Jernej Barbi{\v{c}}} {and}
  \bibinfo{person}{Doug~L James}.} \bibinfo{year}{2005}\natexlab{}.
\newblock \showarticletitle{Real-time subspace integration for St.
  Venant-Kirchhoff deformable models}.
\newblock \bibinfo{journal}{\emph{ACM transactions on graphics (TOG)}}
  \bibinfo{volume}{24}, \bibinfo{number}{3} (\bibinfo{year}{2005}),
  \bibinfo{pages}{982--990}.
\newblock


\bibitem[\protect\citeauthoryear{Barbi{\v{c}}, Sin, and Grinspun}{Barbi{\v{c}}
  et~al\mbox{.}}{2012}]%
        {barbivc2012interactive}
\bibfield{author}{\bibinfo{person}{Jernej Barbi{\v{c}}},
  \bibinfo{person}{Funshing Sin}, {and} \bibinfo{person}{Eitan Grinspun}.}
  \bibinfo{year}{2012}\natexlab{}.
\newblock \showarticletitle{Interactive editing of deformable simulations}.
\newblock \bibinfo{journal}{\emph{ACM Transactions on Graphics (TOG)}}
  \bibinfo{volume}{31}, \bibinfo{number}{4} (\bibinfo{year}{2012}),
  \bibinfo{pages}{70}.
\newblock


\bibitem[\protect\citeauthoryear{Brandt, Eisemann, and Hildebrandt}{Brandt
  et~al\mbox{.}}{2018}]%
        {brandt2018hyper}
\bibfield{author}{\bibinfo{person}{Christopher Brandt}, \bibinfo{person}{Elmar
  Eisemann}, {and} \bibinfo{person}{Klaus Hildebrandt}.}
  \bibinfo{year}{2018}\natexlab{}.
\newblock \showarticletitle{Hyper-reduced projective dynamics}.
\newblock \bibinfo{journal}{\emph{ACM Transactions on Graphics (TOG)}}
  \bibinfo{volume}{37}, \bibinfo{number}{4} (\bibinfo{year}{2018}),
  \bibinfo{pages}{80}.
\newblock


\bibitem[\protect\citeauthoryear{Chen, Cohen-Or, Sorkine, and Toledo}{Chen
  et~al\mbox{.}}{2005}]%
        {chen2005algebraic}
\bibfield{author}{\bibinfo{person}{Doron Chen}, \bibinfo{person}{Daniel
  Cohen-Or}, \bibinfo{person}{Olga Sorkine}, {and} \bibinfo{person}{Sivan
  Toledo}.} \bibinfo{year}{2005}\natexlab{}.
\newblock \showarticletitle{Algebraic analysis of high-pass quantization}.
\newblock \bibinfo{journal}{\emph{ACM Transactions on Graphics (TOG)}}
  \bibinfo{volume}{24}, \bibinfo{number}{4} (\bibinfo{year}{2005}),
  \bibinfo{pages}{1259--1282}.
\newblock


\bibitem[\protect\citeauthoryear{Cong, Bhat, and Fedkiw}{Cong
  et~al\mbox{.}}{2016}]%
        {cong2016art}
\bibfield{author}{\bibinfo{person}{Matthew Cong}, \bibinfo{person}{Kiran~S
  Bhat}, {and} \bibinfo{person}{Ronald Fedkiw}.}
  \bibinfo{year}{2016}\natexlab{}.
\newblock \showarticletitle{Art-directed muscle simulation for high-end facial
  animation}. In \bibinfo{booktitle}{\emph{Symposium on Computer Animation}}.
  \bibinfo{pages}{119--127}.
\newblock


\bibitem[\protect\citeauthoryear{Deng, Chiang, Fox, and Neumann}{Deng
  et~al\mbox{.}}{2006}]%
        {deng2006animating}
\bibfield{author}{\bibinfo{person}{Zhigang Deng}, \bibinfo{person}{Pei-Ying
  Chiang}, \bibinfo{person}{Pamela Fox}, {and} \bibinfo{person}{Ulrich
  Neumann}.} \bibinfo{year}{2006}\natexlab{}.
\newblock \showarticletitle{Animating blendshape faces by cross-mapping motion
  capture data}. In \bibinfo{booktitle}{\emph{Proceedings of the 2006 symposium
  on Interactive 3D graphics and games}}. ACM, \bibinfo{pages}{43--48}.
\newblock


\bibitem[\protect\citeauthoryear{Gao, Yang, Qiao, Lai, Rosin, Xu, and Xia}{Gao
  et~al\mbox{.}}{2018}]%
        {gao2018automatic}
\bibfield{author}{\bibinfo{person}{Lin Gao}, \bibinfo{person}{Jie Yang},
  \bibinfo{person}{Yi-Ling Qiao}, \bibinfo{person}{Yu-Kun Lai},
  \bibinfo{person}{Paul~L Rosin}, \bibinfo{person}{Weiwei Xu}, {and}
  \bibinfo{person}{Shihong Xia}.} \bibinfo{year}{2018}\natexlab{}.
\newblock \showarticletitle{Automatic unpaired shape deformation transfer}. In
  \bibinfo{booktitle}{\emph{SIGGRAPH Asia 2018 Technical Papers}}. ACM,
  \bibinfo{pages}{237}.
\newblock


\bibitem[\protect\citeauthoryear{Isola, Zhu, Zhou, and Efros}{Isola
  et~al\mbox{.}}{2017}]%
        {isola2017image}
\bibfield{author}{\bibinfo{person}{Phillip Isola}, \bibinfo{person}{Jun-Yan
  Zhu}, \bibinfo{person}{Tinghui Zhou}, {and} \bibinfo{person}{Alexei~A
  Efros}.} \bibinfo{year}{2017}\natexlab{}.
\newblock \showarticletitle{Image-to-image translation with conditional
  adversarial networks}. In \bibinfo{booktitle}{\emph{Proceedings of the IEEE
  conference on computer vision and pattern recognition}}.
  \bibinfo{pages}{1125--1134}.
\newblock


\bibitem[\protect\citeauthoryear{Jacobson, Baran, Popovic, and
  Sorkine}{Jacobson et~al\mbox{.}}{2011}]%
        {jacobson2011bounded}
\bibfield{author}{\bibinfo{person}{Alec Jacobson}, \bibinfo{person}{Ilya
  Baran}, \bibinfo{person}{Jovan Popovic}, {and} \bibinfo{person}{Olga
  Sorkine}.} \bibinfo{year}{2011}\natexlab{}.
\newblock \showarticletitle{Bounded biharmonic weights for real-time
  deformation.}
\newblock \bibinfo{journal}{\emph{ACM Trans. Graph.}} \bibinfo{volume}{30},
  \bibinfo{number}{4} (\bibinfo{year}{2011}), \bibinfo{pages}{78}.
\newblock


\bibitem[\protect\citeauthoryear{Joshi, Meyer, DeRose, Green, and
  Sanocki}{Joshi et~al\mbox{.}}{2007}]%
        {joshi2007harmonic}
\bibfield{author}{\bibinfo{person}{Pushkar Joshi}, \bibinfo{person}{Mark
  Meyer}, \bibinfo{person}{Tony DeRose}, \bibinfo{person}{Brian Green}, {and}
  \bibinfo{person}{Tom Sanocki}.} \bibinfo{year}{2007}\natexlab{}.
\newblock \showarticletitle{Harmonic coordinates for character articulation}.
\newblock \bibinfo{journal}{\emph{ACM Transactions on Graphics (TOG)}}
  \bibinfo{volume}{26}, \bibinfo{number}{3} (\bibinfo{year}{2007}),
  \bibinfo{pages}{71}.
\newblock


\bibitem[\protect\citeauthoryear{Joshi, Tien, Desbrun, and Pighin}{Joshi
  et~al\mbox{.}}{2006}]%
        {joshi2006learning}
\bibfield{author}{\bibinfo{person}{Pushkar Joshi}, \bibinfo{person}{Wen~C
  Tien}, \bibinfo{person}{Mathieu Desbrun}, {and}
  \bibinfo{person}{Fr{\'e}d{\'e}ric Pighin}.} \bibinfo{year}{2006}\natexlab{}.
\newblock \showarticletitle{Learning controls for blend shape based realistic
  facial animation}. In \bibinfo{booktitle}{\emph{ACM Siggraph 2006 Courses}}.
  ACM, \bibinfo{pages}{17}.
\newblock


\bibitem[\protect\citeauthoryear{Ju, Schaefer, and Warren}{Ju
  et~al\mbox{.}}{2005}]%
        {ju2005mean}
\bibfield{author}{\bibinfo{person}{Tao Ju}, \bibinfo{person}{Scott Schaefer},
  {and} \bibinfo{person}{Joe Warren}.} \bibinfo{year}{2005}\natexlab{}.
\newblock \showarticletitle{Mean value coordinates for closed triangular
  meshes}.
\newblock \bibinfo{journal}{\emph{ACM Transactions on Graphics (TOG)}}
  \bibinfo{volume}{24}, \bibinfo{number}{3} (\bibinfo{year}{2005}),
  \bibinfo{pages}{561--566}.
\newblock


\bibitem[\protect\citeauthoryear{Kavan, Collins, and O'Sullivan}{Kavan
  et~al\mbox{.}}{2009}]%
        {kavan2009automatic}
\bibfield{author}{\bibinfo{person}{Ladislav Kavan}, \bibinfo{person}{Steven
  Collins}, {and} \bibinfo{person}{Carol O'Sullivan}.}
  \bibinfo{year}{2009}\natexlab{}.
\newblock \showarticletitle{Automatic linearization of nonlinear skinning}. In
  \bibinfo{booktitle}{\emph{Proceedings of the 2009 symposium on Interactive 3D
  graphics and games}}. ACM, \bibinfo{pages}{49--56}.
\newblock


\bibitem[\protect\citeauthoryear{Kavan, Collins, {\v{Z}}{\'a}ra, and
  O'Sullivan}{Kavan et~al\mbox{.}}{2008}]%
        {kavan2008geometric}
\bibfield{author}{\bibinfo{person}{Ladislav Kavan}, \bibinfo{person}{Steven
  Collins}, \bibinfo{person}{Ji{\v{r}}{\'\i} {\v{Z}}{\'a}ra}, {and}
  \bibinfo{person}{Carol O'Sullivan}.} \bibinfo{year}{2008}\natexlab{}.
\newblock \showarticletitle{Geometric skinning with approximate dual quaternion
  blending}.
\newblock \bibinfo{journal}{\emph{ACM Transactions on Graphics (TOG)}}
  \bibinfo{volume}{27}, \bibinfo{number}{4} (\bibinfo{year}{2008}),
  \bibinfo{pages}{105}.
\newblock


\bibitem[\protect\citeauthoryear{Kavan and Sorkine}{Kavan and Sorkine}{2012}]%
        {kavan2012elasticity}
\bibfield{author}{\bibinfo{person}{Ladislav Kavan} {and} \bibinfo{person}{Olga
  Sorkine}.} \bibinfo{year}{2012}\natexlab{}.
\newblock \showarticletitle{Elasticity-inspired deformers for character
  articulation}.
\newblock \bibinfo{journal}{\emph{ACM Transactions on Graphics (TOG)}}
  \bibinfo{volume}{31}, \bibinfo{number}{6} (\bibinfo{year}{2012}),
  \bibinfo{pages}{196}.
\newblock


\bibitem[\protect\citeauthoryear{Kavan and {\v{Z}}{\'a}ra}{Kavan and
  {\v{Z}}{\'a}ra}{2005}]%
        {kavan2005spherical}
\bibfield{author}{\bibinfo{person}{Ladislav Kavan} {and}
  \bibinfo{person}{Ji{\v{r}}{\'\i} {\v{Z}}{\'a}ra}.}
  \bibinfo{year}{2005}\natexlab{}.
\newblock \showarticletitle{Spherical blend skinning: a real-time deformation
  of articulated models}. In \bibinfo{booktitle}{\emph{Proceedings of the 2005
  symposium on Interactive 3D graphics and games}}. ACM,
  \bibinfo{pages}{9--16}.
\newblock


\bibitem[\protect\citeauthoryear{Kim, Pons-Moll, Pujades, Bang, Kim, Black, and
  Lee}{Kim et~al\mbox{.}}{2017}]%
        {kim2017data}
\bibfield{author}{\bibinfo{person}{Meekyoung Kim}, \bibinfo{person}{Gerard
  Pons-Moll}, \bibinfo{person}{Sergi Pujades}, \bibinfo{person}{Seungbae Bang},
  \bibinfo{person}{Jinwook Kim}, \bibinfo{person}{Michael~J Black}, {and}
  \bibinfo{person}{Sung-Hee Lee}.} \bibinfo{year}{2017}\natexlab{}.
\newblock \showarticletitle{Data-driven physics for human soft tissue
  animation}.
\newblock \bibinfo{journal}{\emph{ACM Transactions on Graphics (TOG)}}
  \bibinfo{volume}{36}, \bibinfo{number}{4} (\bibinfo{year}{2017}),
  \bibinfo{pages}{54}.
\newblock


\bibitem[\protect\citeauthoryear{Krysl, Lall, and Marsden}{Krysl
  et~al\mbox{.}}{2001}]%
        {krysl2001dimensional}
\bibfield{author}{\bibinfo{person}{Petr Krysl}, \bibinfo{person}{Sanjay Lall},
  {and} \bibinfo{person}{Jerrold~E Marsden}.} \bibinfo{year}{2001}\natexlab{}.
\newblock \showarticletitle{Dimensional model reduction in non-linear finite
  element dynamics of solids and structures}.
\newblock \bibinfo{journal}{\emph{International Journal for numerical methods
  in engineering}} \bibinfo{volume}{51}, \bibinfo{number}{4}
  (\bibinfo{year}{2001}), \bibinfo{pages}{479--504}.
\newblock


\bibitem[\protect\citeauthoryear{Laine, Karras, Aila, Herva, Saito, Yu, Li, and
  Lehtinen}{Laine et~al\mbox{.}}{2017}]%
        {laine2017production}
\bibfield{author}{\bibinfo{person}{Samuli Laine}, \bibinfo{person}{Tero
  Karras}, \bibinfo{person}{Timo Aila}, \bibinfo{person}{Antti Herva},
  \bibinfo{person}{Shunsuke Saito}, \bibinfo{person}{Ronald Yu},
  \bibinfo{person}{Hao Li}, {and} \bibinfo{person}{Jaakko Lehtinen}.}
  \bibinfo{year}{2017}\natexlab{}.
\newblock \showarticletitle{Production-level facial performance capture using
  deep convolutional neural networks}. In \bibinfo{booktitle}{\emph{Proceedings
  of the ACM SIGGRAPH/Eurographics Symposium on Computer Animation}}. ACM,
  \bibinfo{pages}{10}.
\newblock


\bibitem[\protect\citeauthoryear{Lau, Chai, Xu, and Shum}{Lau
  et~al\mbox{.}}{2009}]%
        {lau2009face}
\bibfield{author}{\bibinfo{person}{Manfred Lau}, \bibinfo{person}{Jinxiang
  Chai}, \bibinfo{person}{Ying-Qing Xu}, {and} \bibinfo{person}{Heung-Yeung
  Shum}.} \bibinfo{year}{2009}\natexlab{}.
\newblock \showarticletitle{Face poser: Interactive modeling of 3d facial
  expressions using facial priors}.
\newblock \bibinfo{journal}{\emph{ACM Transactions on Graphics (TOG)}}
  \bibinfo{volume}{29}, \bibinfo{number}{1} (\bibinfo{year}{2009}),
  \bibinfo{pages}{3}.
\newblock


\bibitem[\protect\citeauthoryear{Le and Hodgins}{Le and Hodgins}{2016}]%
        {le2016real}
\bibfield{author}{\bibinfo{person}{Binh~Huy Le} {and}
  \bibinfo{person}{Jessica~K Hodgins}.} \bibinfo{year}{2016}\natexlab{}.
\newblock \showarticletitle{Real-time skeletal skinning with optimized centers
  of rotation}.
\newblock \bibinfo{journal}{\emph{ACM Transactions on Graphics (TOG)}}
  \bibinfo{volume}{35}, \bibinfo{number}{4} (\bibinfo{year}{2016}),
  \bibinfo{pages}{37}.
\newblock


\bibitem[\protect\citeauthoryear{Le and Lewis}{Le and Lewis}{2019}]%
        {le2019direct}
\bibfield{author}{\bibinfo{person}{Binh~Huy Le} {and} \bibinfo{person}{JP
  Lewis}.} \bibinfo{year}{2019}\natexlab{}.
\newblock \showarticletitle{Direct delta mush skinning and variants}.
\newblock \bibinfo{journal}{\emph{ACM Transactions on Graphics (TOG)}}
  \bibinfo{volume}{38}, \bibinfo{number}{4} (\bibinfo{year}{2019}),
  \bibinfo{pages}{113}.
\newblock


\bibitem[\protect\citeauthoryear{Lewis, Anjyo, Rhee, Zhang, Pighin, and
  Deng}{Lewis et~al\mbox{.}}{2014}]%
        {lewis2014practice}
\bibfield{author}{\bibinfo{person}{John~P Lewis}, \bibinfo{person}{Ken Anjyo},
  \bibinfo{person}{Taehyun Rhee}, \bibinfo{person}{Mengjie Zhang},
  \bibinfo{person}{Frederic~H Pighin}, {and} \bibinfo{person}{Zhigang Deng}.}
  \bibinfo{year}{2014}\natexlab{}.
\newblock \showarticletitle{Practice and Theory of Blendshape Facial Models.}
\newblock \bibinfo{journal}{\emph{Eurographics (State of the Art Reports)}}
  \bibinfo{volume}{1}, \bibinfo{number}{8} (\bibinfo{year}{2014}),
  \bibinfo{pages}{2}.
\newblock


\bibitem[\protect\citeauthoryear{Lewis and Anjyo}{Lewis and Anjyo}{2010}]%
        {lewis2010direct}
\bibfield{author}{\bibinfo{person}{John~P Lewis} {and}
  \bibinfo{person}{Ken-ichi Anjyo}.} \bibinfo{year}{2010}\natexlab{}.
\newblock \showarticletitle{Direct manipulation blendshapes}.
\newblock \bibinfo{journal}{\emph{IEEE Computer Graphics and Applications}}
  \bibinfo{volume}{30}, \bibinfo{number}{4} (\bibinfo{year}{2010}),
  \bibinfo{pages}{42--50}.
\newblock


\bibitem[\protect\citeauthoryear{Lewis, Cordner, and Fong}{Lewis
  et~al\mbox{.}}{2000}]%
        {lewis2000pose}
\bibfield{author}{\bibinfo{person}{John~P Lewis}, \bibinfo{person}{Matt
  Cordner}, {and} \bibinfo{person}{Nickson Fong}.}
  \bibinfo{year}{2000}\natexlab{}.
\newblock \showarticletitle{Pose space deformation: a unified approach to shape
  interpolation and skeleton-driven deformation}. In
  \bibinfo{booktitle}{\emph{Proceedings of the 27th annual conference on
  Computer graphics and interactive techniques}}. ACM Press/Addison-Wesley
  Publishing Co., \bibinfo{pages}{165--172}.
\newblock


\bibitem[\protect\citeauthoryear{Li, Weise, and Pauly}{Li
  et~al\mbox{.}}{2010}]%
        {li2010example}
\bibfield{author}{\bibinfo{person}{Hao Li}, \bibinfo{person}{Thibaut Weise},
  {and} \bibinfo{person}{Mark Pauly}.} \bibinfo{year}{2010}\natexlab{}.
\newblock \showarticletitle{Example-based facial rigging}. In
  \bibinfo{booktitle}{\emph{Acm transactions on graphics (tog)}},
  Vol.~\bibinfo{volume}{29}. ACM, \bibinfo{pages}{32}.
\newblock


\bibitem[\protect\citeauthoryear{Lipman, Levin, and Cohen-Or}{Lipman
  et~al\mbox{.}}{2008}]%
        {lipman2008green}
\bibfield{author}{\bibinfo{person}{Yaron Lipman}, \bibinfo{person}{David
  Levin}, {and} \bibinfo{person}{Daniel Cohen-Or}.}
  \bibinfo{year}{2008}\natexlab{}.
\newblock \showarticletitle{Green coordinates}.
\newblock \bibinfo{journal}{\emph{ACM Transactions on Graphics (TOG)}}
  \bibinfo{volume}{27}, \bibinfo{number}{3} (\bibinfo{year}{2008}),
  \bibinfo{pages}{78}.
\newblock


\bibitem[\protect\citeauthoryear{Liu, Zheng, Tang, Yuan, Fan, and Zhou}{Liu
  et~al\mbox{.}}{2019}]%
        {liu2019neuroskinning}
\bibfield{author}{\bibinfo{person}{Lijuan Liu}, \bibinfo{person}{Youyi Zheng},
  \bibinfo{person}{Di Tang}, \bibinfo{person}{Yi Yuan},
  \bibinfo{person}{Changjie Fan}, {and} \bibinfo{person}{Kun Zhou}.}
  \bibinfo{year}{2019}\natexlab{}.
\newblock \showarticletitle{NeuroSkinning: automatic skin binding for
  production characters with deep graph networks}.
\newblock \bibinfo{journal}{\emph{ACM Transactions on Graphics (TOG)}}
  \bibinfo{volume}{38}, \bibinfo{number}{4} (\bibinfo{year}{2019}),
  \bibinfo{pages}{114}.
\newblock


\bibitem[\protect\citeauthoryear{Loper, Mahmood, Romero, Pons-Moll, and
  Black}{Loper et~al\mbox{.}}{2015}]%
        {loper2015smpl}
\bibfield{author}{\bibinfo{person}{Matthew Loper}, \bibinfo{person}{Naureen
  Mahmood}, \bibinfo{person}{Javier Romero}, \bibinfo{person}{Gerard
  Pons-Moll}, {and} \bibinfo{person}{Michael~J Black}.}
  \bibinfo{year}{2015}\natexlab{}.
\newblock \showarticletitle{SMPL: A skinned multi-person linear model}.
\newblock \bibinfo{journal}{\emph{ACM transactions on graphics (TOG)}}
  \bibinfo{volume}{34}, \bibinfo{number}{6} (\bibinfo{year}{2015}),
  \bibinfo{pages}{248}.
\newblock


\bibitem[\protect\citeauthoryear{Luo, Shao, Wang, Xu, Chen, Zhou, and Yang}{Luo
  et~al\mbox{.}}{2018}]%
        {luo2018nnwarp}
\bibfield{author}{\bibinfo{person}{Ran Luo}, \bibinfo{person}{Tianjia Shao},
  \bibinfo{person}{Huamin Wang}, \bibinfo{person}{Weiwei Xu},
  \bibinfo{person}{Xiang Chen}, \bibinfo{person}{Kun Zhou}, {and}
  \bibinfo{person}{Yin Yang}.} \bibinfo{year}{2018}\natexlab{}.
\newblock \showarticletitle{NNWarp: Neural Network-based Nonlinear
  Deformation}.
\newblock \bibinfo{journal}{\emph{IEEE transactions on visualization and
  computer graphics}} (\bibinfo{year}{2018}).
\newblock


\bibitem[\protect\citeauthoryear{Magnenat-Thalmann, Laperrire, and
  Thalmann}{Magnenat-Thalmann et~al\mbox{.}}{1988}]%
        {magnenat1988joint}
\bibfield{author}{\bibinfo{person}{Nadia Magnenat-Thalmann},
  \bibinfo{person}{Richard Laperrire}, {and} \bibinfo{person}{Daniel
  Thalmann}.} \bibinfo{year}{1988}\natexlab{}.
\newblock \showarticletitle{Joint-dependent local deformations for hand
  animation and object grasping}. In \bibinfo{booktitle}{\emph{In Proceedings
  on Graphics interface’88}}. Citeseer.
\newblock


\bibitem[\protect\citeauthoryear{Mancewicz, Derksen, Rijpkema, and
  Wilson}{Mancewicz et~al\mbox{.}}{2014}]%
        {mancewicz2014delta}
\bibfield{author}{\bibinfo{person}{Joe Mancewicz}, \bibinfo{person}{Matt~L
  Derksen}, \bibinfo{person}{Hans Rijpkema}, {and} \bibinfo{person}{Cyrus~A
  Wilson}.} \bibinfo{year}{2014}\natexlab{}.
\newblock \showarticletitle{Delta Mush: smoothing deformations while preserving
  detail}. In \bibinfo{booktitle}{\emph{Proceedings of the Fourth Symposium on
  Digital Production}}. ACM, \bibinfo{pages}{7--11}.
\newblock


\bibitem[\protect\citeauthoryear{Merry, Marais, and Gain}{Merry
  et~al\mbox{.}}{2006}]%
        {merry2006animation}
\bibfield{author}{\bibinfo{person}{Bruce Merry}, \bibinfo{person}{Patrick
  Marais}, {and} \bibinfo{person}{James Gain}.}
  \bibinfo{year}{2006}\natexlab{}.
\newblock \showarticletitle{Animation space: A truly linear framework for
  character animation}.
\newblock \bibinfo{journal}{\emph{ACM Transactions on Graphics (TOG)}}
  \bibinfo{volume}{25}, \bibinfo{number}{4} (\bibinfo{year}{2006}),
  \bibinfo{pages}{1400--1423}.
\newblock


\bibitem[\protect\citeauthoryear{Meyer and Anderson}{Meyer and
  Anderson}{2007}]%
        {meyer2007key}
\bibfield{author}{\bibinfo{person}{Mark Meyer} {and} \bibinfo{person}{John
  Anderson}.} \bibinfo{year}{2007}\natexlab{}.
\newblock \showarticletitle{Key point subspace acceleration and soft caching}.
\newblock \bibinfo{journal}{\emph{ACM Transactions on Graphics (TOG)}}
  \bibinfo{volume}{26}, \bibinfo{number}{3} (\bibinfo{year}{2007}),
  \bibinfo{pages}{74}.
\newblock


\bibitem[\protect\citeauthoryear{Mukai}{Mukai}{2015}]%
        {mukai2015building}
\bibfield{author}{\bibinfo{person}{Tomohiko Mukai}.}
  \bibinfo{year}{2015}\natexlab{}.
\newblock \showarticletitle{Building helper bone rigs from examples}. In
  \bibinfo{booktitle}{\emph{Proceedings of the 19th Symposium on Interactive 3D
  Graphics and Games}}. ACM, \bibinfo{pages}{77--84}.
\newblock


\bibitem[\protect\citeauthoryear{Mukai and Kuriyama}{Mukai and
  Kuriyama}{2016}]%
        {mukai2016efficient}
\bibfield{author}{\bibinfo{person}{Tomohiko Mukai} {and}
  \bibinfo{person}{Shigeru Kuriyama}.} \bibinfo{year}{2016}\natexlab{}.
\newblock \showarticletitle{Efficient dynamic skinning with low-rank helper
  bone controllers}.
\newblock \bibinfo{journal}{\emph{ACM Transactions on Graphics (TOG)}}
  \bibinfo{volume}{35}, \bibinfo{number}{4} (\bibinfo{year}{2016}),
  \bibinfo{pages}{36}.
\newblock


\bibitem[\protect\citeauthoryear{Pentland and Williams}{Pentland and
  Williams}{1989}]%
        {pentland1989good}
\bibfield{author}{\bibinfo{person}{Alexander Pentland} {and}
  \bibinfo{person}{John Williams}.} \bibinfo{year}{1989}\natexlab{}.
\newblock \showarticletitle{Good vibrations: Modal dynamics for graphics and
  animation}.
\newblock  (\bibinfo{year}{1989}).
\newblock


\bibitem[\protect\citeauthoryear{Si, Lee, Sifakis, and Terzopoulos}{Si
  et~al\mbox{.}}{2014}]%
        {si2014realistic}
\bibfield{author}{\bibinfo{person}{Weiguang Si}, \bibinfo{person}{Sung-Hee
  Lee}, \bibinfo{person}{Eftychios Sifakis}, {and} \bibinfo{person}{Demetri
  Terzopoulos}.} \bibinfo{year}{2014}\natexlab{}.
\newblock \showarticletitle{Realistic biomechanical simulation and control of
  human swimming}.
\newblock \bibinfo{journal}{\emph{ACM Transactions on Graphics (TOG)}}
  \bibinfo{volume}{34}, \bibinfo{number}{1} (\bibinfo{year}{2014}),
  \bibinfo{pages}{10}.
\newblock


\bibitem[\protect\citeauthoryear{Sloan, Rose~III, and Cohen}{Sloan
  et~al\mbox{.}}{2001}]%
        {sloan2001shape}
\bibfield{author}{\bibinfo{person}{Peter-Pike~J Sloan},
  \bibinfo{person}{Charles~F Rose~III}, {and} \bibinfo{person}{Michael~F
  Cohen}.} \bibinfo{year}{2001}\natexlab{}.
\newblock \showarticletitle{Shape by example}. In
  \bibinfo{booktitle}{\emph{Proceedings of the 2001 symposium on Interactive 3D
  graphics}}. ACM, \bibinfo{pages}{135--143}.
\newblock


\bibitem[\protect\citeauthoryear{Sorkine}{Sorkine}{2005}]%
        {sorkine2005laplacian}
\bibfield{author}{\bibinfo{person}{Olga Sorkine}.}
  \bibinfo{year}{2005}\natexlab{}.
\newblock \showarticletitle{Laplacian mesh processing}. In
  \bibinfo{booktitle}{\emph{Eurographics (STARs)}}. \bibinfo{pages}{53--70}.
\newblock


\bibitem[\protect\citeauthoryear{Sorkine and Alexa}{Sorkine and Alexa}{2007}]%
        {sorkine2007rigid}
\bibfield{author}{\bibinfo{person}{Olga Sorkine} {and} \bibinfo{person}{Marc
  Alexa}.} \bibinfo{year}{2007}\natexlab{}.
\newblock \showarticletitle{As-rigid-as-possible surface modeling}. In
  \bibinfo{booktitle}{\emph{Symposium on Geometry processing}},
  Vol.~\bibinfo{volume}{4}. \bibinfo{pages}{109--116}.
\newblock


\bibitem[\protect\citeauthoryear{Sorkine, Cohen-Or, Irony, and Toledo}{Sorkine
  et~al\mbox{.}}{2005}]%
        {sorkine2005geometry}
\bibfield{author}{\bibinfo{person}{Olga Sorkine}, \bibinfo{person}{Daniel
  Cohen-Or}, \bibinfo{person}{Dror Irony}, {and} \bibinfo{person}{Sivan
  Toledo}.} \bibinfo{year}{2005}\natexlab{}.
\newblock \showarticletitle{Geometry-aware bases for shape approximation}.
\newblock \bibinfo{journal}{\emph{IEEE transactions on visualization and
  computer graphics}} \bibinfo{volume}{11}, \bibinfo{number}{2}
  (\bibinfo{year}{2005}), \bibinfo{pages}{171--180}.
\newblock


\bibitem[\protect\citeauthoryear{Sumner, Schmid, and Pauly}{Sumner
  et~al\mbox{.}}{2007}]%
        {sumner2007embedded}
\bibfield{author}{\bibinfo{person}{Robert~W Sumner}, \bibinfo{person}{Johannes
  Schmid}, {and} \bibinfo{person}{Mark Pauly}.}
  \bibinfo{year}{2007}\natexlab{}.
\newblock \showarticletitle{Embedded deformation for shape manipulation}.
\newblock In \bibinfo{booktitle}{\emph{ACM SIGGRAPH 2007 papers}}.
  \bibinfo{pages}{80--es}.
\newblock


\bibitem[\protect\citeauthoryear{Tan, Gao, Lai, and Xia}{Tan
  et~al\mbox{.}}{2018a}]%
        {tan2018variational}
\bibfield{author}{\bibinfo{person}{Qingyang Tan}, \bibinfo{person}{Lin Gao},
  \bibinfo{person}{Yu-Kun Lai}, {and} \bibinfo{person}{Shihong Xia}.}
  \bibinfo{year}{2018}\natexlab{a}.
\newblock \showarticletitle{Variational autoencoders for deforming 3d mesh
  models}. In \bibinfo{booktitle}{\emph{Proceedings of the IEEE Conference on
  Computer Vision and Pattern Recognition}}. \bibinfo{pages}{5841--5850}.
\newblock


\bibitem[\protect\citeauthoryear{Tan, Gao, Lai, Yang, and Xia}{Tan
  et~al\mbox{.}}{2018b}]%
        {tan2018mesh}
\bibfield{author}{\bibinfo{person}{Qingyang Tan}, \bibinfo{person}{Lin Gao},
  \bibinfo{person}{Yu-Kun Lai}, \bibinfo{person}{Jie Yang}, {and}
  \bibinfo{person}{Shihong Xia}.} \bibinfo{year}{2018}\natexlab{b}.
\newblock \showarticletitle{Mesh-based autoencoders for localized deformation
  component analysis}. In \bibinfo{booktitle}{\emph{Thirty-Second AAAI
  Conference on Artificial Intelligence}}.
\newblock


\bibitem[\protect\citeauthoryear{Wang, Pulli, and Popovi{\'c}}{Wang
  et~al\mbox{.}}{2007}]%
        {wang2007real}
\bibfield{author}{\bibinfo{person}{Robert~Y Wang}, \bibinfo{person}{Kari
  Pulli}, {and} \bibinfo{person}{Jovan Popovi{\'c}}.}
  \bibinfo{year}{2007}\natexlab{}.
\newblock \showarticletitle{Real-time enveloping with rotational regression}.
  In \bibinfo{booktitle}{\emph{ACM Transactions on Graphics (TOG)}},
  Vol.~\bibinfo{volume}{26}. ACM, \bibinfo{pages}{73}.
\newblock


\bibitem[\protect\citeauthoryear{Wang and Phillips}{Wang and Phillips}{2002}]%
        {wang2002multi}
\bibfield{author}{\bibinfo{person}{Xiaohuan~Corina Wang} {and}
  \bibinfo{person}{Cary Phillips}.} \bibinfo{year}{2002}\natexlab{}.
\newblock \showarticletitle{Multi-weight enveloping: least-squares
  approximation techniques for skin animation}. In
  \bibinfo{booktitle}{\emph{Proceedings of the 2002 ACM SIGGRAPH/Eurographics
  symposium on Computer animation}}. ACM, \bibinfo{pages}{129--138}.
\newblock


\bibitem[\protect\citeauthoryear{Wang, Jacobson, Barbi{\v{c}}, and Kavan}{Wang
  et~al\mbox{.}}{2015}]%
        {wang2015linear}
\bibfield{author}{\bibinfo{person}{Yu Wang}, \bibinfo{person}{Alec Jacobson},
  \bibinfo{person}{Jernej Barbi{\v{c}}}, {and} \bibinfo{person}{Ladislav
  Kavan}.} \bibinfo{year}{2015}\natexlab{}.
\newblock \showarticletitle{Linear subspace design for real-time shape
  deformation}.
\newblock \bibinfo{journal}{\emph{ACM Transactions on Graphics (TOG)}}
  \bibinfo{volume}{34}, \bibinfo{number}{4} (\bibinfo{year}{2015}),
  \bibinfo{pages}{1--11}.
\newblock


\bibitem[\protect\citeauthoryear{Weise, Bouaziz, Li, and Pauly}{Weise
  et~al\mbox{.}}{2011}]%
        {weise2011realtime}
\bibfield{author}{\bibinfo{person}{Thibaut Weise}, \bibinfo{person}{Sofien
  Bouaziz}, \bibinfo{person}{Hao Li}, {and} \bibinfo{person}{Mark Pauly}.}
  \bibinfo{year}{2011}\natexlab{}.
\newblock \showarticletitle{Realtime performance-based facial animation}. In
  \bibinfo{booktitle}{\emph{ACM transactions on graphics (TOG)}},
  Vol.~\bibinfo{volume}{30}. ACM, \bibinfo{pages}{77}.
\newblock


\bibitem[\protect\citeauthoryear{Zhang, Van~Kaick, and Dyer}{Zhang
  et~al\mbox{.}}{2010}]%
        {zhang2010spectral}
\bibfield{author}{\bibinfo{person}{Hao Zhang}, \bibinfo{person}{Oliver
  Van~Kaick}, {and} \bibinfo{person}{Ramsay Dyer}.}
  \bibinfo{year}{2010}\natexlab{}.
\newblock \showarticletitle{Spectral mesh processing}. In
  \bibinfo{booktitle}{\emph{Computer graphics forum}},
  Vol.~\bibinfo{volume}{29}. Wiley Online Library, \bibinfo{pages}{1865--1894}.
\newblock


\end{thebibliography}

\end{document}